\documentclass[amsmath,amssymb,pre]{revtex4}

\usepackage{graphicx}
\usepackage{dcolumn}
\usepackage{subfigure}
\usepackage{latexsym,amssymb,amsmath}
\usepackage{bm}

\begin{document}


\title{Granular packings with moving side walls}

\author{James W. Landry}
 \email{jwlandr@sandia.gov}
\author{Gary S. Grest}
 \affiliation{Sandia National Laboratories, Albuquerque, New Mexico 87185}

\date{June 25, 2003}

\begin{abstract}
The effects of movement of the side walls of a confined granular packing
are studied by discrete element, molecular dynamics simulations.  The
dynamical evolution of the stress is studied as a function of wall
movement both in the direction of gravity as well as opposite to it.
For all wall velocities explored, the stress in the final state of the
system after wall movement is fundamentally different from the original
state obtained by pouring particles into the container and letting them
settle under the influence of gravity.  The original packing possesses a
hydrostatic-like region at the top of the container which crosses over
to a depth-independent stress.  As the walls are moved in the direction
opposite to gravity, the saturation stress first reaches a minimum value
independent of the wall velocity, then increases to a steady-state value
dependent on the wall-velocity.  After wall movement ceases and the
packing reaches equilibrium, the stress profile fits the classic Janssen
form for high wall velocities, while it has some deviations for low wall
velocities.  The wall movement greatly increases the number of
particle-wall and particle-particle forces at the Coulomb criterion.
Varying the wall velocity has only small effects on the particle
structure of the final packing so long as the walls travel a similar
distance.
\end{abstract}

\maketitle

\section{\label{sec:introduction}Introduction}

There has been a recent resurgence in interest in the formation and
structure of granular packings in the physics
community~\cite{JaegerOct1996,Duran2000}.  One particular facet of
granular packings to receive attention recently is their stress
profiles, both because of the industrial applications of
silos~\cite{Nedderman1992}, and because of new experimental techniques
to measure effective stress in packings~\cite{MuethMar1998,
VanelMay1999, VanelFeb2000}.  The first theoretical attempt to
understand stress in a silo geometry dates back over 100 years to
Janssen~\cite{Janssen1895}, who obtained a one-parameter form for the
vertical stress in a silo.  Several assumptions were made to arrive at
this result.  One was to treat the granular material as a continuous
medium where a fraction $\kappa$ of vertical stress is converted to
horizontal stress.  Another assumption was that the forces of friction
between particles and walls are at the Coulomb failure criterion: $F_t =
\mu_w F_n$, where $F_t$ is the magnitude of the tangential friction
force, $F_n$ is the normal force at the wall, and $\mu_w$ is the
coefficient of friction for particle-wall contacts.  This assumption is
also known as incipient failure.  For a cylindrical container of radius
$R$ with static wall friction $\mu_w$ and granular pack of total height
$z_0$, the Janssen analysis predicts the vertical stress
$\sigma_{zz}(z)$ at a height $z$ is
\begin{equation}
\sigma_{zz}(z) = \rho g l\left[1 -
\exp\left(-\frac{z_0 - z}{l}\right)\right]
\label{eq:janssen}
\end{equation}
where the decay length $l = \frac{R}{2\kappa\mu_w}$. $\kappa$ represents
the fraction of the weight carried by the side walls, $\rho$ is the
volumetric density, $g$ is gravity, and $z_0$ is height of the top of
the packing.

Numerous experiments have been carried out to verify this theory, but
precise experiments are difficult.  Recently, extremely well-controlled
experiments on granular packings have been
done~\cite{VanelMay1999,VanelFeb2000} that show a deviation from the
ideal Janssen form, which we shall hereafter refer to as the
Vanel-Cl\'{e}ment form.  This phenomenological form with one additional
free parameter includes a hydrostatic-like region of linear dependence
of pressure with depth (defined by the length $a$), followed by a region
that conforms to the Janssen theory:
\begin{eqnarray}
z_0 - z < a & : & \sigma_{zz}(z) = \rho g (z_0 - z) \nonumber\\ 
z_0 - z > a & : & \sigma_{zz}(z) = \rho g \left(a + l\left[1 - \exp\left(-\frac{z_0 - z - a}{l}\right)\right]\right)
\label{eq:twoparm}
\end{eqnarray}
This form was also found in extensive molecular dynamics simulations of
granular packings in both two and three
dimensions~\cite{LandryApr2003,Landry0302115}.  These packings were
created both through pouring and sedimentation and then allowed to
settle under the influence of gravity.

Many questions about stress in granular packings still remain
unanswered.  Even after a packing has been formed, it may be perturbed
in many ways that radically change its stress profile and physical
structure.  Many studies have focused on tapping as a means to compress
the packing and its logarithmic response
time~\cite{KnightMay1995,NowakFeb1998,JosserandOct2000,PhilippeDec2002}.
Another method to perturb a packing is to move the side
walls~\cite{BerthoApr2003,OvarlezJun2003,OvarlezSep2003}.  The effect of
this movement is not well understood and is the focus of this study.

Recently, experiments have been conducted on granular packings in
cylindrical containers with movable side
walls~\cite{BerthoApr2003,OvarlezJun2003,OvarlezSep2003}.  The
experiments make use of a movable cylinder enclosing a granular packing
supported by an independent base.  These experiments find over a wide
range of wall velocities good agreement with the Janssen form for the
vertical stress after the system has relaxed following cessation of wall
movement.  This is in contrast to the earlier experiments on packings
with fixed side walls~\cite{VanelFeb2000,VanelMay1999}.  Here we present
large-scale three-dimensional ($3D$) discrete element, molecular
dynamics simulations of granular packings in cylindrical containers
(silos) with movable side walls.  Our aim is to understand how the
motion of the side wall modifies the stress profiles in granular
packings and to compare our results with the recent experimental
findings~\cite{BerthoApr2003,OvarlezJun2003,OvarlezSep2003}.  We analyze
how wall movement and its cessation affect the stress profile of the
packings.  We also investigate in depth the effects of wall movement on
the internal structure and particle positions of these packings, which
cannot be easily measured experimentally.  The behavior of the system
under wall movement is similar across a wide range of wall velocities.
Finally we show that wall movement in the direction opposite to gravity
drives tangential forces to the Coulomb criterion everywhere, leading
directly to the Janssen form for the stress profiles.

The simulation technique and model are presented in Sec. II.  The stress
profiles and their features are discussed in Sec. III.  In Sec. IV we
examine the particle motion during and after wall movement, while Sec. V
we discuss the force distribution of the resultant packings.  A brief
summary and conclusions are presented in Sec. VI.

\section{\label{sec:simulation}Simulation Method}

We present discrete element, molecular dynamics (MD) simulations in $3D$
of model systems of $N=50 000$ mono-dispersed spheres of fixed mass $m$
and diameter $d$.  The system is constrained by a cylinder of radius
$R=10d$, centered on $x=y=0$, with its axis along the vertical $z$
direction.  The cylinder is bounded below by a layer of
randomly-arranged immobilized particles approximately $2d$ high to
provide a rough base.  This work builds on previous MD simulations of
packings in silos, where further details of the model can be
found~\cite{LandryApr2003}.

The spheres interact only on contact through a spring-dashpot
interaction in the normal and tangential directions to their lines of
centers.  Contacting spheres $i$ and $j$ positioned at $\mathbf{r}_i$
and $\mathbf{r}_j$ experience a relative normal compression $\delta =
|\mathbf{r}_{ij} - d|$, where $\mathbf{r}_{ij} = \mathbf{r}_i -
\mathbf{r}_j$, which results in a force
\begin{equation}
\mathbf{F}_{ij} = \mathbf{F}_n + \mathbf{F}_t.
\end{equation}
The normal and tangential contact forces are given by 
\begin{equation}
\mathbf{F}_{n} = f(\delta/d) (k_n \delta {\mathbf n}_{ij} - \frac{m}{2} \gamma_n \mathbf{v}_n)
\end{equation}
\begin{equation}
\mathbf{F}_{t} = f(\delta/d) (-k_t \mathbf{\Delta s}_t - \frac{m}{2} \gamma_t \mathbf{v}_t)
\end{equation}
where $\mathbf{n}_{ij} = \mathbf{r}_{ij}/r_{ij}$, with $r_{ij} =
|\mathbf{r}_{ij}|$. $\mathbf{v}_n$ and $\mathbf{v}_t$ are the normal and
tangential components of the relative surface velocity, and $k_{n,t}$
and $\gamma_{n,t}$ are elastic and viscoelastic constants,
respectively. $f(x) = 1$ for Hookean (linear) contacts while for
Hertzian contacts $f(x) = \sqrt{x}$.  $\mathbf{\Delta s}_t$ is the
elastic tangential displacement between spheres, obtained by integrating
tangential relative velocities during elastic deformation for the
lifetime of the contact.  The magnitude of $\mathbf{\Delta s}_t$ is
truncated as necessary to satisfy a local Coulomb yield criterion $F_t
\le \mu F_n$, where $F_t \equiv |\mathbf{F}_t|$ and $F_n \equiv
|\mathbf{F}_n|$ and $\mu$ is the particle-particle friction coefficient.
Frictionless spheres correspond to $\mu = 0$.  Particle-wall
interactions are treated the same, but the particle-wall friction
coefficient $\mu_w$ is set independently.  The side wall of the
container is smooth, and thus the particle-wall normal force is always
perpendicular to the $xy$ plane.  A more detailed description of the
model is available elsewhere~\cite{SilbertOct2001,LandryApr2003}.

These simulations are run with a fixed set of parameters: $k_n = 2
\times 10^5 mg/d$, $k_t = \frac{2}{7} k_n$, and $\gamma_n =
50\sqrt{g/d}$.  For Hookean springs we set $\gamma_t = 0$.  In these
simulations, it takes far longer to drain the energy out of granular
packs using the Hertzian force law, since the coefficient of restitution
$\epsilon$ is velocity-dependent~\cite{SchaeferJan1996} and goes to zero
as the velocity goes to zero.  We thus use Hookean
contacts~\cite{HertzianNote1}, which for the above parameters give
$\epsilon = 0.88$.  The convenient time unit is $\tau = \sqrt{d/g}$, the
time it takes a particle to fall its radius from rest under gravity.
For this set of parameters, the timestep $\delta t = 10^{-4} \tau$.  The
particle-particle friction and particle-wall friction are the same: $\mu
= \mu_w = 0.5$.  The simulations are run using a parallel distributed
memory code on 20 DEC Alpha processors.  One million timesteps takes
approximately 7 hours.  Our longest simulation for slow wall velocities,
$v = 5 \times 10^{-4}\, d/\tau$, ran for $t = 1.14 \times 10^{4}\,\tau$,
which corresponds to approximately 800 hours.

The simulations all begin with the same initial packing to minimize
sample to sample fluctuations.  This packing was generated by pouring
particles into a container using method P1 as described in
ref.~\cite{LandryApr2003}.  The beginning packing is quiescent.  Over
the course of a simulation, the cylindrical side wall of the packing
moves for a time $t_s$, which usually is $10^3 \tau$, or over a fixed
distance $\Delta z$.  After this period, the walls cease to move and the
packing settles.  We consider a packing quiescent when the kinetic
energy per particle $E_k \le 10^{-8} mgd$.  The time scale for this
relaxation is very short, usually less than $10 \tau$. The cylindrical
side wall is moved either up ($+z$) or down ($-z$) with a constant
velocity $v_s$ varying from $10^{-1}$ to $10^{-5} d/\tau$.  As in the
experiments~\cite{BerthoApr2003,OvarlezJun2003,OvarlezSep2003}, only the
side wall moves --- the rough base is immobile throughout the course of
the simulation.

The wall velocities used here for upward velocities are in the range
used in the experiments by Bertho \textit{et al.}~\cite{BerthoApr2003}.
In that study, glass beads with $d = 2\,\mathrm{mm}$ were moved with a
velocity $v_s$ ranging from $2 \times 10^{-2}$ to $35\,\mathrm{mm/s}$ or
$1.4 \times 10^{-4}$ to $0.25\,d/\tau$ over distances up to $\Delta z =
70\,\mathrm{mm} = 35d$.  Ovarlez \textit{et al.}~\cite{OvarlezJun2003}
use glass beads with $d = 1.5$ mm and a fixed $v_s = 1.5 \times 10^{-3}$
mm/s or $1.2 \times 10^{-5}\,d/\tau$ over a distance $\Delta z \approx
1.5 \times 10^{-2}\,\mathrm{mm} = 10^{-2} d$.  This is a very low
velocity over a very short distance.  In an earlier
study~\cite{VanelFeb2000,VanelMay1999}, Vanel \textit{et al.} used a
higher velocity, in conjunction with tapping, of $v_s = 2 \times
10^{-2}\,\mathrm{mm/s}$ for particles of the same diameter.  Another
distinction between the experiments by Ovarlez et
al.~\cite{OvarlezJun2003} and Vanel et
al.~\cite{VanelFeb2000,VanelMay1999} was the time at which measurement
occurs.  In the Ovarlez \textit{et al.} experiments, measurement occurs
right at the end of wall movement, while in the Vanel \textit{et al.}
experiments, the packing is allowed to settle before measurements are
made and tapping is sometimes also applied.  Ovarlez and Cl\'{e}ment
also studied~\cite{OvarlezSep2003} moving the wall downwards with a
$v_s$ ranging from $-5\,\mathrm{nm/s}$ to $-100\,\mu\mathrm{m/s}$, which
corresponds to wall velocities of $-4 \times 10^{-8}\,d/\tau$ to $-8
\times 10^{-3}\,d/\tau$.  This range of velocities overlaps with our
velocity range for downward wall movement, but also extends to much
slower velocities.

\begin{figure*}
\includegraphics[width=1.0in,clip]{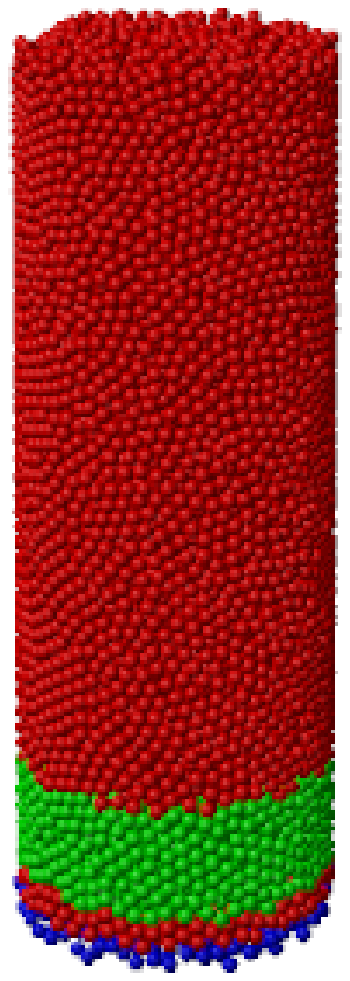}
\hspace{0.5in}
\includegraphics[width=1.0in,clip]{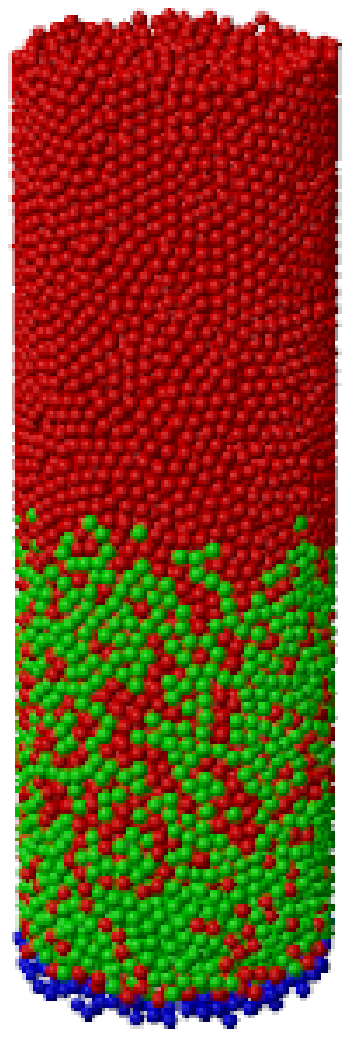}
\hspace{0.5in}
\includegraphics[width=1.0in,clip]{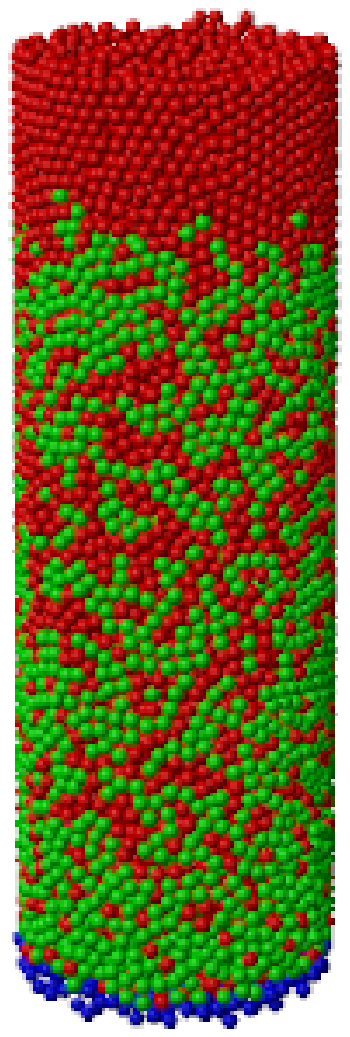}
\hspace{0.5in}
\includegraphics[width=1.0in,clip]{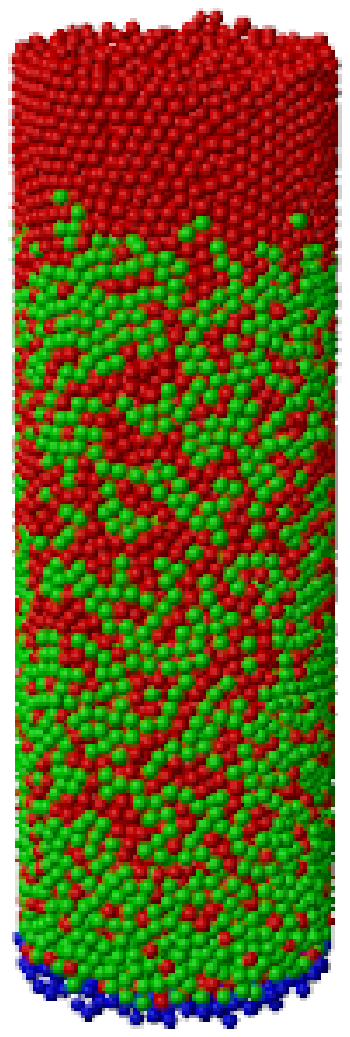}
\caption{\label{fig:move}(color online) Effects of wall movement on a $N
= 20 000$ packing.  The particles are shown in dark gray (red) and the
fixed particles that form the base are shown in black (blue).  All
particles with initial positions between $z=5$ and $z=15$ have been
colored light gray (green). The cylindrical wall is moved upwards ($+z$)
with a velocity $v_s = 10^{-1} d/\tau$, a very high velocity.  (a) the
starting configuration before the wall moves.  (b) the packing after the
wall has moved for $5 \times 10^2 \tau$, $\Delta z = 50d$.  The
packing is now taller - it has fluffed up as the wall has moved.
Significant particle rearrangement is occurring.  (c) the packing after
the wall has moved for $10^3 \tau$, $\Delta z = 100d$.  Particles in
contact with the wall have been dragged significantly up the pile.  (d)
The wall has stopped after $10^3 \tau$ and the packing has reached
equilibrium.  There is little particle movement during relaxation.}
\end{figure*}

Packings are examined both during wall movement and after cessation of
wall movement and settling.  Figure~\ref{fig:move} shows the structure
of a packing for a smaller system of $20000$ particles (used for
illustration) with wall velocity $v_s = 10^{-1} d/\tau$ in the upward
($+z$) direction.  This is a very high velocity, so during wall
movement, there is significant particle rearrangement, and a number of
particles originally in contact with the wall move upward.  Particles
between $z=5$ and $z=15$ have been colored light gray (green online) to
provide a visual picture of particle movement over time.  The height of
the pile changes significantly during wall movement (4.7 \%), but does
not change after the wall movement has ceased and the packing reaches
equilibrium.  For much slower wall velocities applied for the same time,
very little particle movement is observed, and the height of the packing
does not change.

\section{\label{sec:stress}Stress Profiles}

In this section we discuss the effect of wall movement on stress
profiles of packings.  In general, wall movement increases the
tangential force between particles and the wall.  The wall moves in
relation to the particles, so the integrated displacement $\Delta s$
increases as the wall moves.  Since the tangential force is proportional
to $\Delta s$, $F_t$ increases over time.  Wall movement thus drives
$F_t$ towards the Coulomb limit, $F_t = \mu_w F_n$, for particles in
contact with the wall.  We know from previous work~\cite{LandryApr2003}
that the prime factor determining the form of the stress profile in
these packings is the tangential force between particle and wall.  If
$F_t$ is close to the Coulomb criterion, then we expect the stress to
follow the Janssen form throughout the pile.  In addition, the more
particle-wall interactions that are at the Coulomb criterion, the
stronger this effect and the lower the value of the saturation stress in
the packing, i.e. the value of the stress in the depths of the packing,
where the stress becomes depth-independent.  We believe this effect is
the origin of the experimental measurements observing remarkable
agreement with the Janssen theory after the side walls of a packing are
moved~\cite{BerthoApr2003,OvarlezJun2003}.  Below, we investigate the
specific effects of wall movement on the stress profile.

\begin{figure}
\includegraphics[width=2.25in,angle=270,clip]{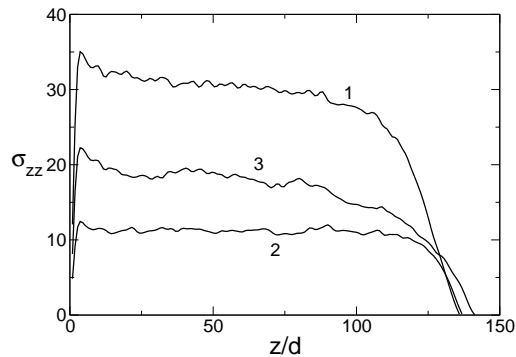}
\caption{\label{fig:wdu-progress}Vertical stress $\sigma_{zz}$ in units
of $mg/d^2$ during wall movement with large velocity $v_s = 0.01
d/\tau$.  Profile 1 corresponds to the packing at $t=0$, the initial
packing.  Profile 2 is the same packing after the wall has moved for $t=
100 \tau$, a distance $\Delta z = d$.  The saturation stress has fallen
to $11 mg/d^2$.  Profile 3 is the same packing at $t = 700 \tau$
($\Delta z = 7d$).  The saturation stress first decreases to a minimum
value and then increases to a steady state.}
\end{figure}

Figure~\ref{fig:wdu-progress} shows the change in the stress profile of
a packing with wall velocity $v_s = 0.01 d/\tau$, a relatively large
velocity.  The stress profile with the largest saturation stress
corresponds to $t=0$, before wall movement has begun.  This stress
profile obeys the Vanel-Cl\'{e}ment
form~\cite{VanelMay1999,LandryApr2003}, with a linear stress profile at
the top of the pile crossing over to a Janssen form in the depth of the
pile.  After moving the wall for $t= 100 \tau$, the saturation stress
has decreased by more than a factor of three to $\sigma_{zz} \simeq
11\,mg/d^2$ and the height of the pile has increased.  As the wall
movement continues, the saturation stress slowly increases.  After a
long time $t \gtrsim 700 \tau$, the system reaches steady state and the
stress profile ceases to change markedly.  Over the course of the wall
movement, the height of the pile has increased by approximately 3\%, and
there has been considerable particle rearrangement.  In addition, the
linear stress disappears almost immediately after the initiation of wall
movement.  This change in the saturation stress is consistent with that
observed in granular experiments performed by Bertho \textit{et
al.}~\cite{BerthoApr2003}, where the apparent mass (effectively the
saturation stress) dropped quickly after the initiation of wall movement
and then slowly increased with time until it reached a saturation value.

\begin{figure}
\includegraphics[width=2.25in,angle=270,clip]{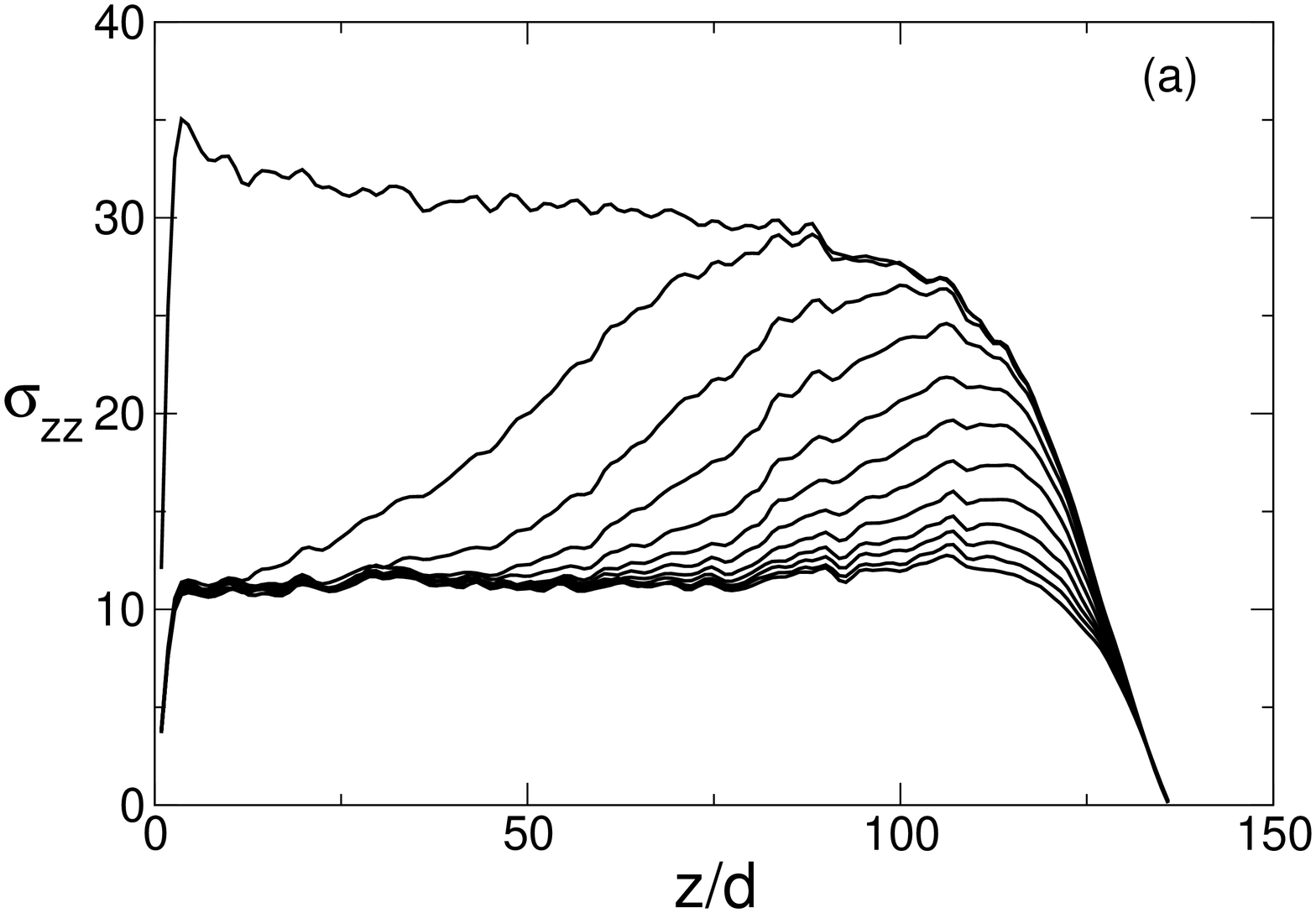}
\includegraphics[width=2.25in,angle=270,clip]{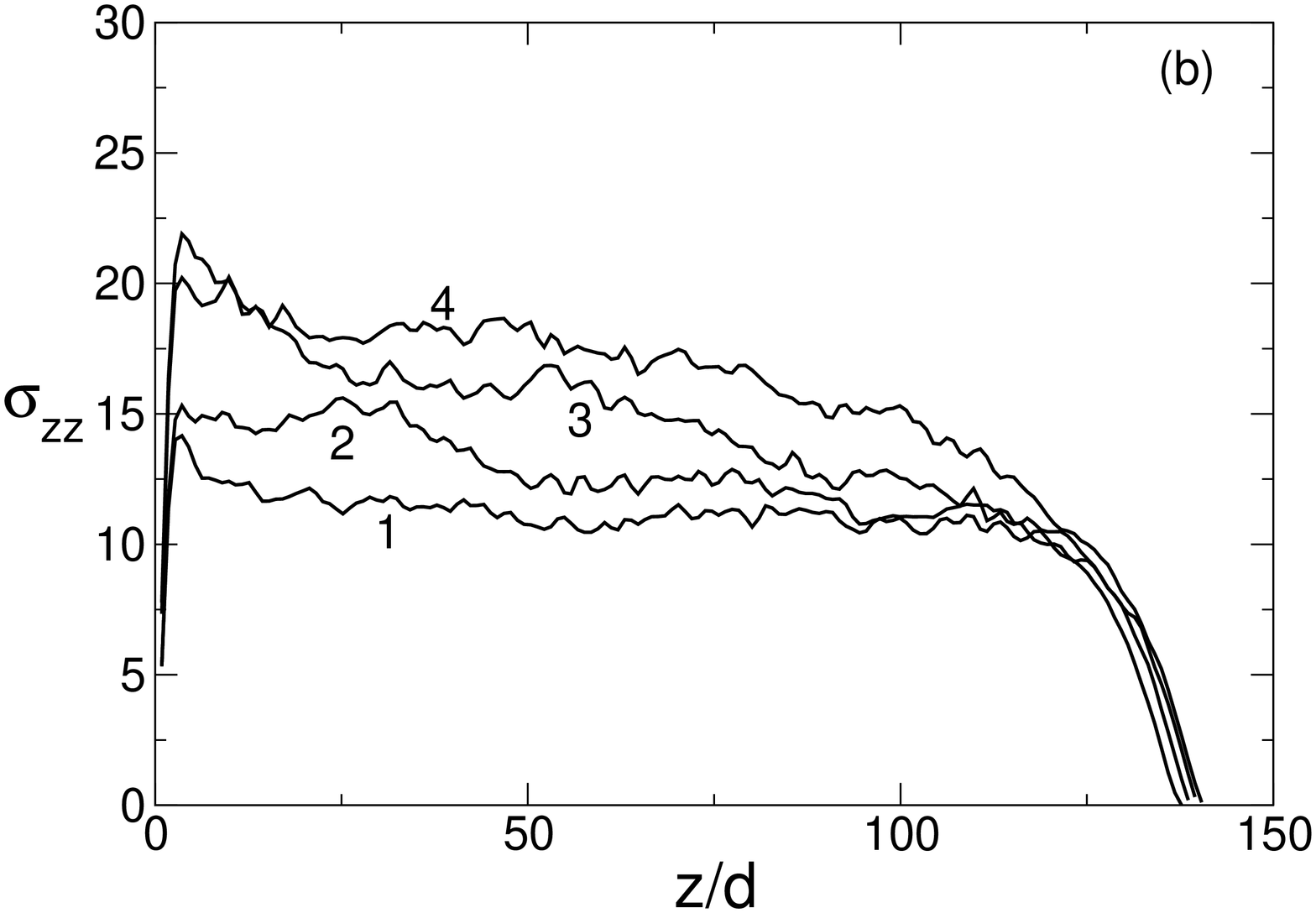}
\caption{\label{fig:wdu-slow}Vertical stress $\sigma_{zz}$ in units of
$mg/d^2$.  (a) Stress progression for wall velocity $v_s = 10^{-4}
\,d/\tau$.  Here each stress profile is $200 \tau$ after the previous
one, with the first profile corresponding to the packing before wall
movement.  The wall velocity is slow enough that the stress profile is
only incrementally disrupted over time and does not attain the Janssen
form until $t = 2200 \tau$ and $\Delta z = 0.22 d$.  (b) Stress profiles
for wall velocity $v_s = 5 \times 10^{-4}\,d/\tau$.  Profile 1 is the
minimum saturation stress $\sigma_{zz} \simeq 11 mg/d^2$ shown at $t =
2.8 \times 10^3 \tau$, $\Delta z = 1.6d$.  Profile 2 is after $t = 6
\times 10^3 \tau$, $\Delta z = 3.1d$; profile 3 $t = 9 \times 10^3
\tau$, $\Delta z = 4.6d$; and profile 4 $t = 1.14 \times 10^4 \tau$,
$\Delta z = 5.7d$.  Over time the height of the packing as well as the
saturation stress increase.}
\end{figure}

The behavior of the stress profile is similar for low wall velocities,
although not identical.  Unfortunately, computational limitations are
much more severe for these packings than for high wall velocity
packings, because the computational time required for the side wall to
travel the same distance is so much larger.  Figure~\ref{fig:wdu-slow}a
shows the change in the stress profile for a relatively slow wall
velocity $v_s = 10^{-4} d/\tau$.  Each profile is $2 \times 10^2 \tau$
after the previous one, starting at $t = 0$.  For this wall velocity,
the stress drops quickly at the base of the pile, but initially remains
unchanged at the top of the pile.  As wall movement continues, the new
reduced stress profile propagates up the pile.  Eventually, after moving
the walls for $t = 2.2 \times 10^3 \tau$, which corresponds to a very
small vertical distance $\Delta z = 0.22d$, the entire stress profile
follows the Janssen form.  At this point in the simulation, the height
of the pile has not changed.  This minimum saturation stress is the same
as that observed in Figure~\ref{fig:wdu-progress}, $\sigma_{zz} \simeq
11\,mg/d^2$, and is stable for a significant period of subsequent wall
movement.  This suggests that this is the limit for low saturation
stress in this packing, and its value is controlled purely by geometric
factors.

We explored the full behavior of the stress profile during wall movement
using a low wall velocity of $v_s = 5 \times 10^{-4} d/\tau$ for $1.14
\times 10^4 \tau$, so that the side walls moved a total distance of
$\Delta z = 5.7d$, which is comparable to the distance walls were moved
for higher velocities.  As shown in Figure~\ref{fig:wdu-slow}b,
prolonged wall movement at low velocities does eventually move the
stress profile away from the minimum saturation stress.  Prolonged wall
movement also increases the height of the pile by 3.3 \%, which is
similar to the height change for packings with high wall velocities.  As
the wall moves upward, changes in the stress profile propagate upwards
from the bottom of the pack to the top.  This occurs not only at early
times, when the minimum saturation stress is propagated up the pile, but
also at later times, when enhanced stress is also propagated up the pile
as further wall movement increases the eventual saturation stress.  This
drop to a minimum saturation stress followed by an increase to a larger,
stable saturation stress is the same for all observed wall velocities.
The drop to a minimum saturation stress occurs at wall movement
distances $\Delta z \approx 0.3 d$ for all wall velocities measured.  In
addition, the increase in saturation stress from the minimum occurs when
the wall has moved a distance $\Delta z \approx 1.6d$.  At this
distance, particles near the wall have completely moved past nearby
particles, which means that whatever contacts these particles had
initially have been destroyed.  The minimum saturation stress is thus
the optimized stress network for a given initial condition, and wall
movement of greater than $d$ destroys the initial stress network, and
forms another, which is no longer optimal.  This process is not steady,
however.  The stress changes relatively abruptly at certain times, then
remains essentially unchanged for long periods (on order of $3 \times
10^3 \tau$ for $v_s = 5 \times 10^{-4} d/\tau$), then abruptly changes
again.  One exception to this is that the height of the packing
increases smoothly with time until it reaches a stable height at $\sim 9
\times 10^3 \tau$ of $140.5d$, after which it remains unchanged under
further wall movement.

\begin{figure}
\includegraphics[width=2.25in,angle=270,clip]{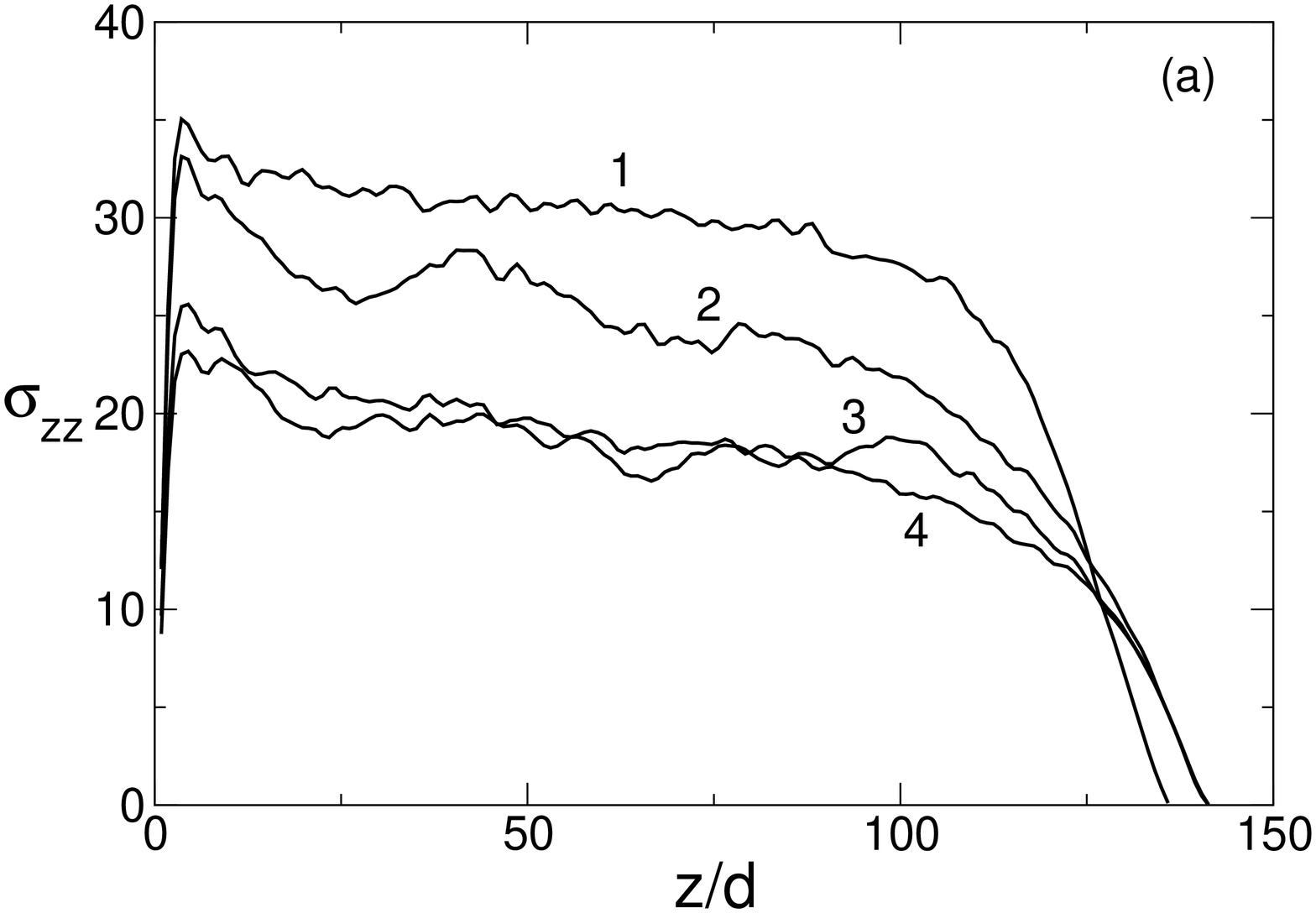}
\includegraphics[width=2.25in,angle=270,clip]{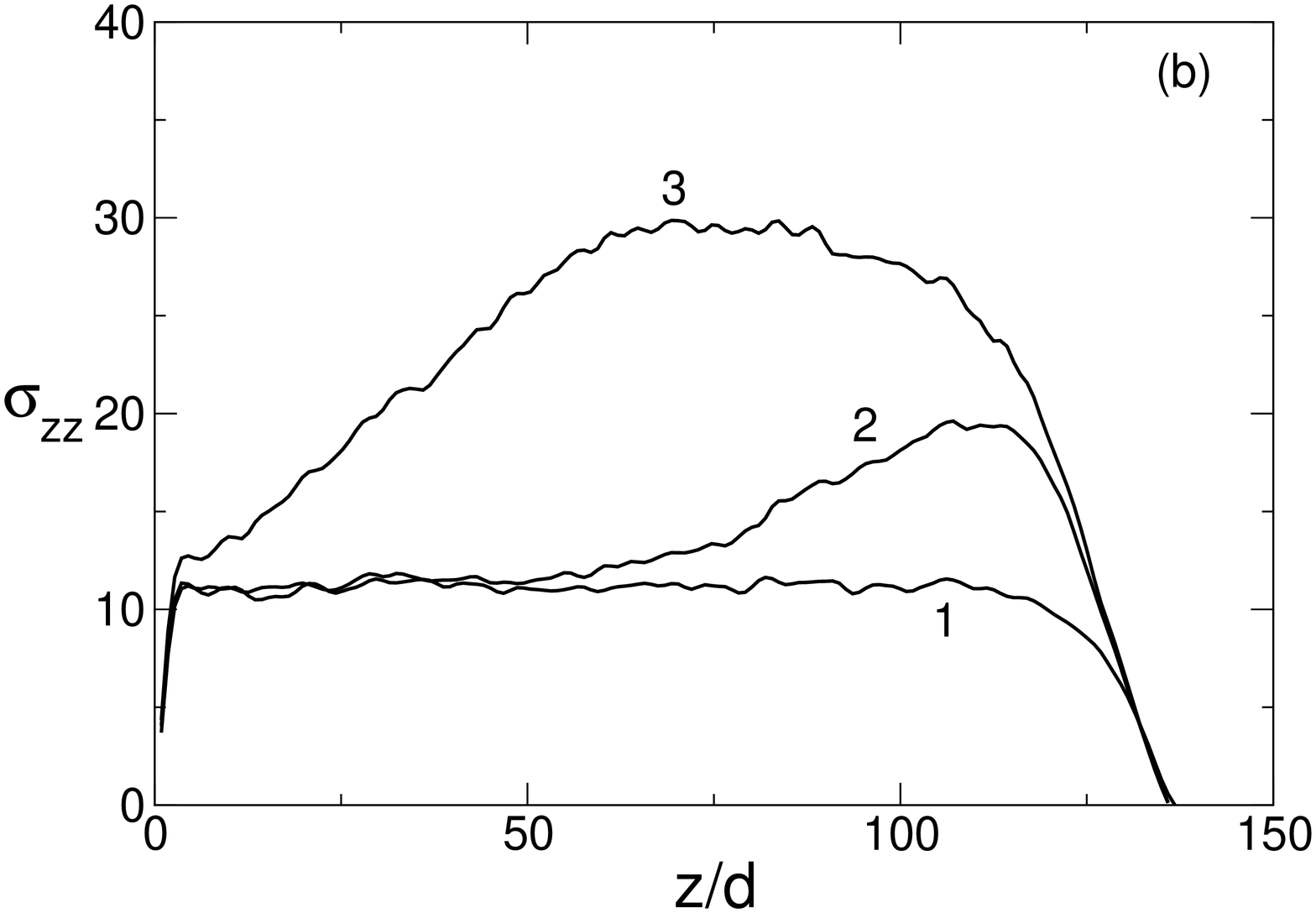}
\caption{\label{fig:wdu-relax}Vertical stress $\sigma_{zz}$ in units of
$mg/d^2$ for packings moved for $t= 10^3 \tau$ and then allowed to
relax.  (a) High wall velocity profiles.  Profile 1 corresponds to the
initial packing.  Profile 2 is the packing for $v_s = 10^{-1} d/\tau$,
profile 3 $v_s = 10^{-2} d/\tau$, and profile 4 $v_s = 10^{-3} d/\tau$.
(b) Low wall velocity profiles.  Profile 1 corresponds to $v_s = 5
\times 10^{-4} d/\tau$, profile 2 to $v_s = 10^{-4} d/\tau$, and profile 3
to $v_s = 10^{-5} d/\tau$.}
\end{figure}

Figure~\ref{fig:wdu-relax} shows the final stress profiles for the same
packings after moving the wall for $t = 10^{3} \tau$ for high and low
wall velocity and then allowing them to settle.  The wall movement has
very different effects on the final stress profile depending on its
strength.  As seen in Figure~\ref{fig:wdu-relax}a, wall velocities of
$v_s \gtrsim 10^{-3} d/\tau$ increase the height of the pile relative to
the original packing, even after relaxation.  The change in height is
substantial and similar for many different wall velocities, 3.7\%,
meaning there has been a large change in the density of the packing.
The final stress after relaxation is related to the magnitude of the
velocity $v_s$, for $v_s \gtrsim 10^{-3} d/\tau$.  Those packings with
larger $v_s$ had larger saturation stress in the final packing.  As we
shall see in Section IV, the larger $v_s$, the more the particles
rearrange.  In addition, the larger $v_s$, the larger the particle
rearrangement after cessation of wall movement is.  This suggests that
particle rearrangement strongly influences the final saturation stress.
Those packings with large $v_s$ have enough particle rearrangement to
completely disrupt the force network.  The higher the velocity, the more
contacts both at the wall and in the bulk are broken and the larger the
eventual saturation stress.  This behavior was also observed by Bertho
\textit{et al.}~\cite{BerthoApr2003}: after the wall stopped moving, the
packing settled and the apparent mass increased.  In addition, they
observed the same trend after relaxation.  The larger the $v_s$, the
larger the final saturation stress.

By contrast, small wall velocities ($v_s < 10^{-3} d/\tau$) do not
change the height of the pile over the same time period, largely because
the wall does not travel far enough to substantially disrupt the force
network.  Large-scale rearrangements do not occur for this time
duration.  In this case, the particles against the wall are fully
mobilized and the resultant saturation stress is small, because most of
the pressure is supported by the walls.  When the packing is allowed to
relax, there are no large particle rearrangements, because the wall
movement is not large enough to move particle positions significantly,
and the wall movement has not put very much energy into the packing.
The packing stays in the minimum saturation stress configuration.  In
addition, if the wall does not move for long enough to force the entire
packing into the minimum saturation stress configuration, the
intermediate stress configuration is stable under relaxation, as shown
in Fig.~\ref{fig:wdu-relax}b for the packing with wall velocity $v_s =
10^{-4} d/\tau$.

If the walls are moved long enough at a low velocity to approach the
stable final saturation stress (higher than the minimum saturation
stress) the height of the pile does increase, as shown in
Figure~\ref{fig:wdu-slow}b for $v_s = 5 \times 10^{-4}\,d/\tau$.  After
relaxation, however, the stress remains unchanged.  The slow wall
movement does not introduce enough energy into the packing for the
stress to change during relaxation, even though for these long times,
there is significant particle rearrangement before relaxation.

\begin{figure}
\includegraphics[width=2.25in,angle=270,clip]{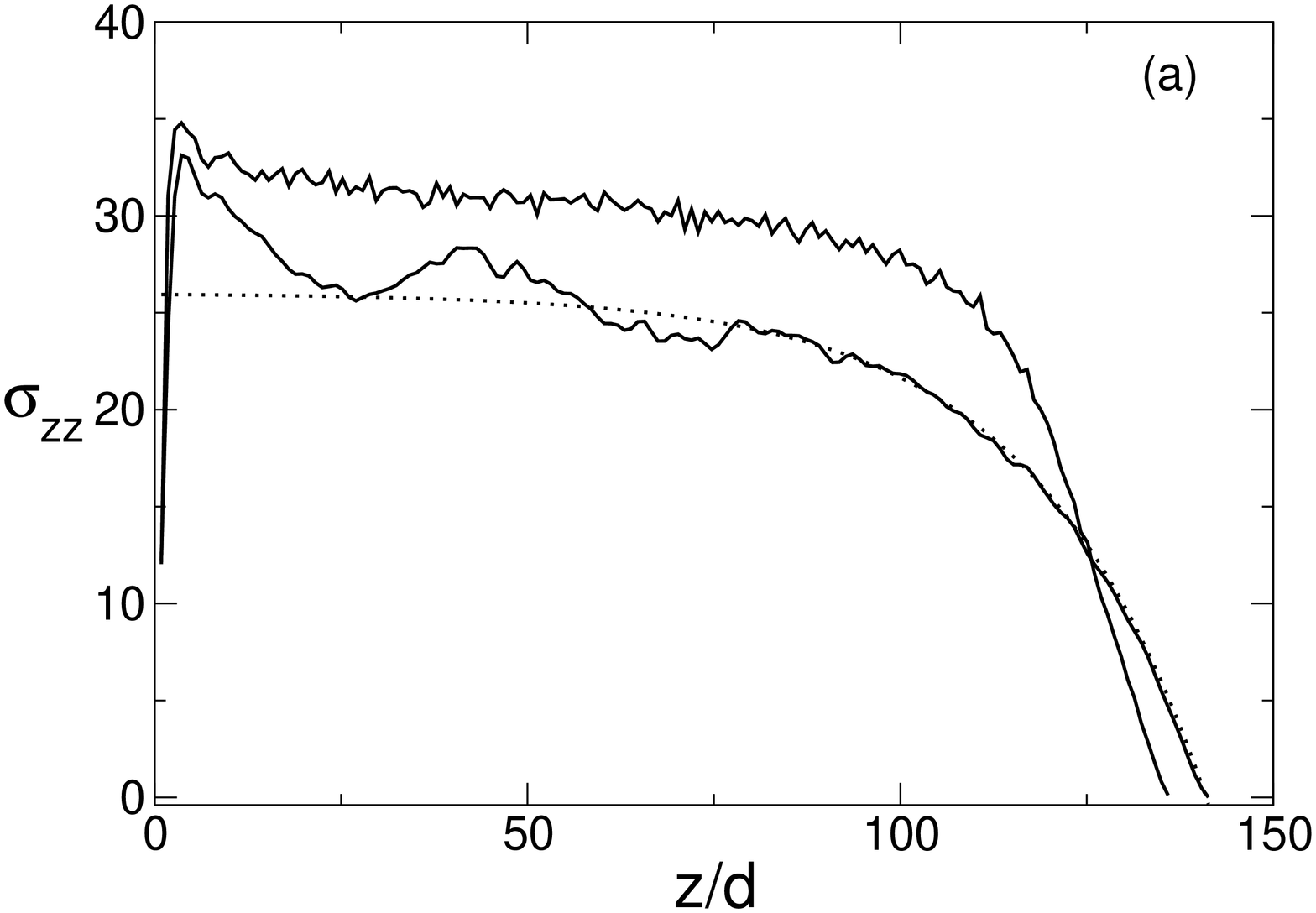}
\includegraphics[width=2.25in,angle=270,clip]{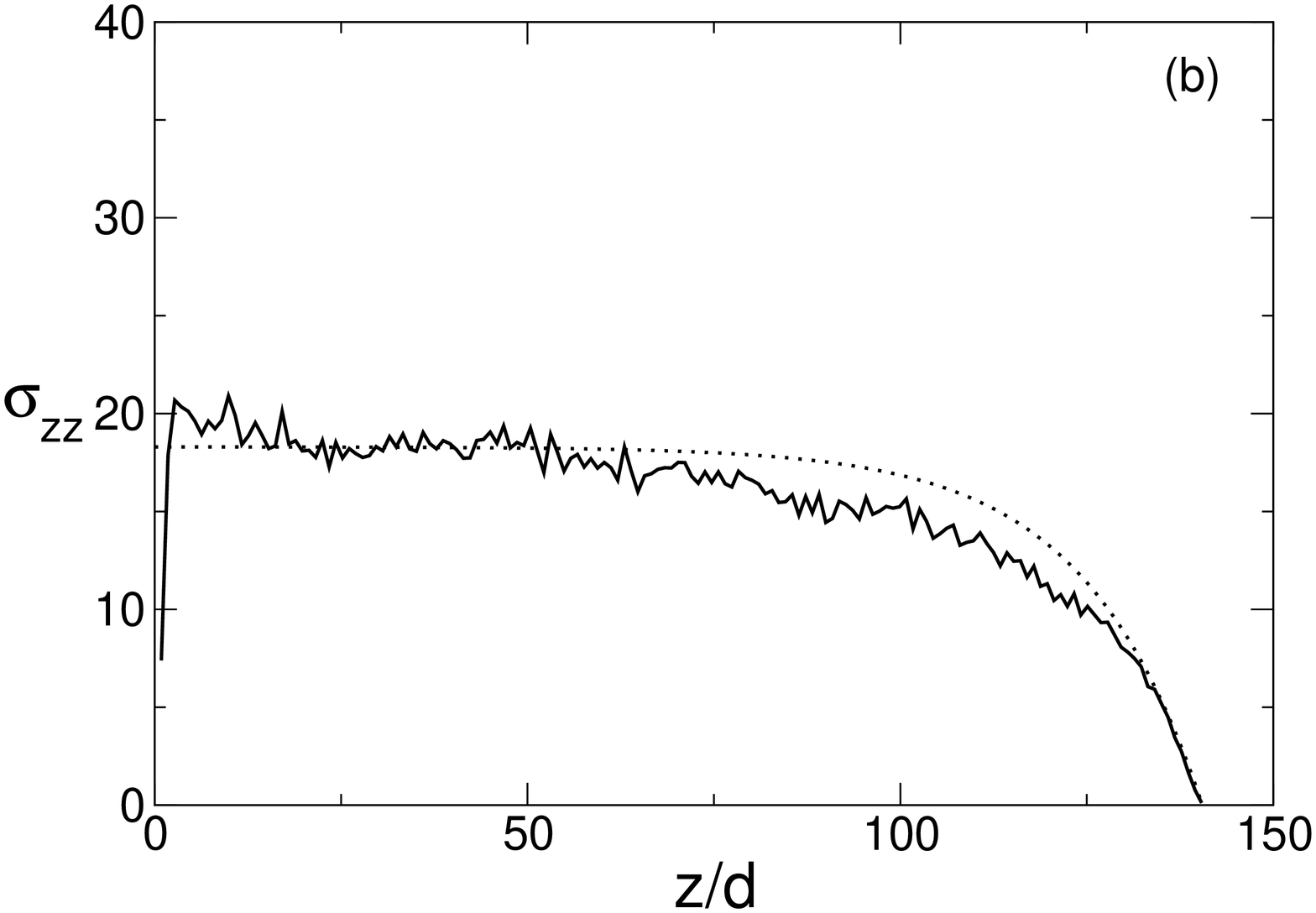}
\caption{\label{fig:wdu-janssen}Vertical stress $\sigma_{zz}$ in units
of $mg/d^2$ for a $N = 50 000$ particle packing.  (a) wall velocity $v_s
= 0.1 d/\tau$ for $\Delta z = 100d$ after cessation and relaxation
compared to the original packing and (b) wall velocity $v_s = 5 \times
10^{-4}\,d/\tau$ for $\Delta z = 5.7d$ after cessation and relaxation.
The fit to the Janssen form is shown as a dotted line.}
\end{figure}

Figure~\ref{fig:wdu-janssen} shows a comparison of the stress profile of
two distinct packings.  The first is the original quiescent packing.
The second packing is this same packing after the wall has moved for $t
= 10^{3} \tau$ at $v_s = 0.1 d/\tau$ and then the packing has settled.
The nature of the stress profile has been radically changed.  While the
original packing fits very well the Vanel-Cl\'{e}ment
form~\cite{VanelMay1999}, the same packing after wall movement fits the
one parameter Janssen form~\cite{Janssen1895} well with $\kappa = 0.44$.
As we shall see in Section V, the wall movement has forced the
tangential force of the particles at the wall to the Coulomb criteria
everywhere, eliminating the linear stress region.  Another example is
given for a slow velocity $v_s = 5 \times 10^{-4}\,d/\tau$, where
$\kappa = 0.63$.  In this case, however, the Janssen fit overshoots the
actual stress near the top of the pack, the opposite of the case where
it fails for the original stress region.  For all velocities observed,
the linear stress region is destroyed, but only the high velocity cases
fit the Janssen stress form well.

\begin{figure}
\includegraphics[width=2.25in,angle=270,clip]{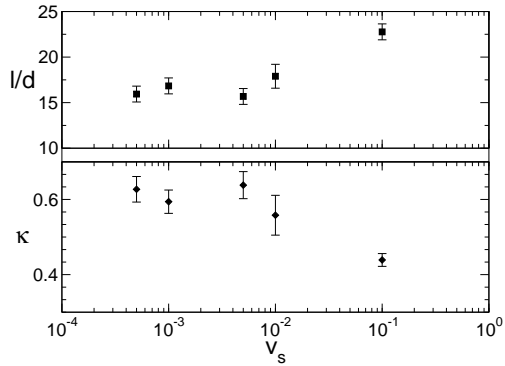}
\caption{\label{fig:kappa} Janssen length $l$ and fraction of weight
  $\kappa$ for different wall velocities after cessation of wall
  movement and relaxation.  Fast wall velocities $v_s \ge
  10^{-3}\,d/\tau$ were applied for $t= 10^{3}\,\tau$, while the $v_s =
  5 \times 10^{-4}\,d/\tau$ data were obtained after applying the wall
  velocity for $t = 1.14 \times 10^{4}\,\tau$.}
\end{figure}

We have compiled $\kappa$'s for the various wall movement rates and show
them in Figure~\ref{fig:kappa}.  These values can be compared with
Figure 4 in Bertho \textit{et al.}~\cite{BerthoApr2003}.  In both cases,
as the wall velocity increases, the resultant Janssen length $l$
increases.  Since $\kappa = R/2 \mu_w l$, $\kappa$ decreases with
increasing wall velocity $v_s$.  The $\kappa$'s observed are much lower
than those obtained for the original packings with the Vanel-Cl\'{e}ment
form, which were slightly greater than 1~\cite{LandryApr2003}.  In our
case, the intermediate stresses between the minimal stress and final
stress do not follow the Janssen form and $\kappa$ for those stress
profiles has no meaning.  Using the observed $l$ and $R=7.5d$ given in
ref.~\cite{BerthoApr2003}, we find $\kappa$'s ranging from $\kappa =
0.349$ for $v_s = 20\,mm/s = 0.14 \,d/\tau$ to $\kappa = 0.405$ for $v_s
= 0.2\,mm/s = 1.4 \times 10^{-3} \,d/\tau$.  These values are slightly
lower but close to our observed values, shown in Fig.~\ref{fig:kappa}.

\begin{figure}
\includegraphics[width=2.25in,angle=270,clip]{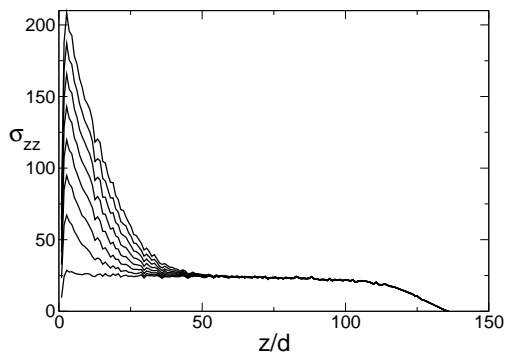}
\caption{\label{fig:wdd}Vertical stress $\sigma_{zz}$ in units of
$mg/d^2$ for a particle packing with wall velocity $v_s = -10^{-4}
d/\tau$ starting at $t = 0$ and every $10^2 \tau$ afterward.  The
stress at the bottom of the packing increases with time.  This stress
does not dissipate after relaxation.}
\end{figure}

Downward motion of the wall is a very different phenomenon.  In this
case, the wall movement merely increases the stress at the base of the
pile, without changing the stress profile elsewhere, as shown in
Figure~\ref{fig:wdd}, which shows the stress profile for a particle
packing with $v_s = -10^{4} d/\tau$ applied for $t_s = 10^3 \tau$.  This
stress buildup increases with duration of wall movement and occurs for
all the velocities probed, from $v_s = -10^{-1} d/\tau$ to $-10^{-5}
d/\tau$.  In addition, this stress buildup is robust and does not
disappear when the packing settles after cessation of wall movement.
The extra stress is locked in.  There is a slight difference between
high and low wall velocities.  Low wall velocities do not change the
height of the packing, while high wall velocities ($v_s \ge
10^{-3}\,d/\tau$) increase the density at the bottom of the pile and
lower the overall height of the packing by a slight amount.

\section{Particle Rearrangement}

We have studied the motion of particles during and after wall movement
to understand the effects of wall movement on the position and density
of the packing.  As the walls are moved upward with velocity $v_s$, they
drag particles in contact with the walls upwards by means of the
frictional force between them.  As particles on the edge of the
cylindrical container move upward, particles in the middle of the
packing move downward and outward toward the walls to fill the voids
created by the upward-moving particles.

\begin{figure}
\includegraphics[width=2.5in,clip]{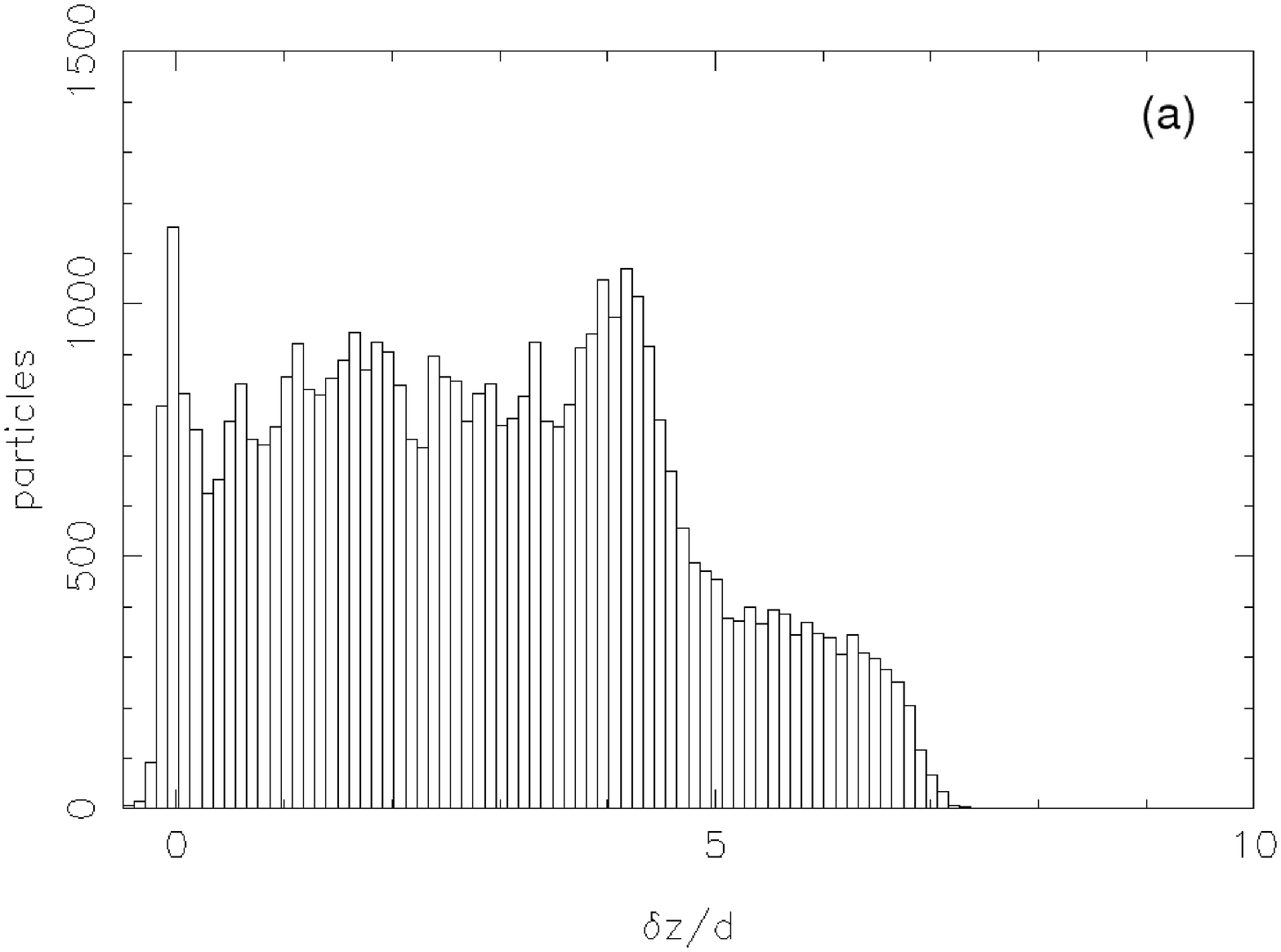}
\includegraphics[width=2in,clip]{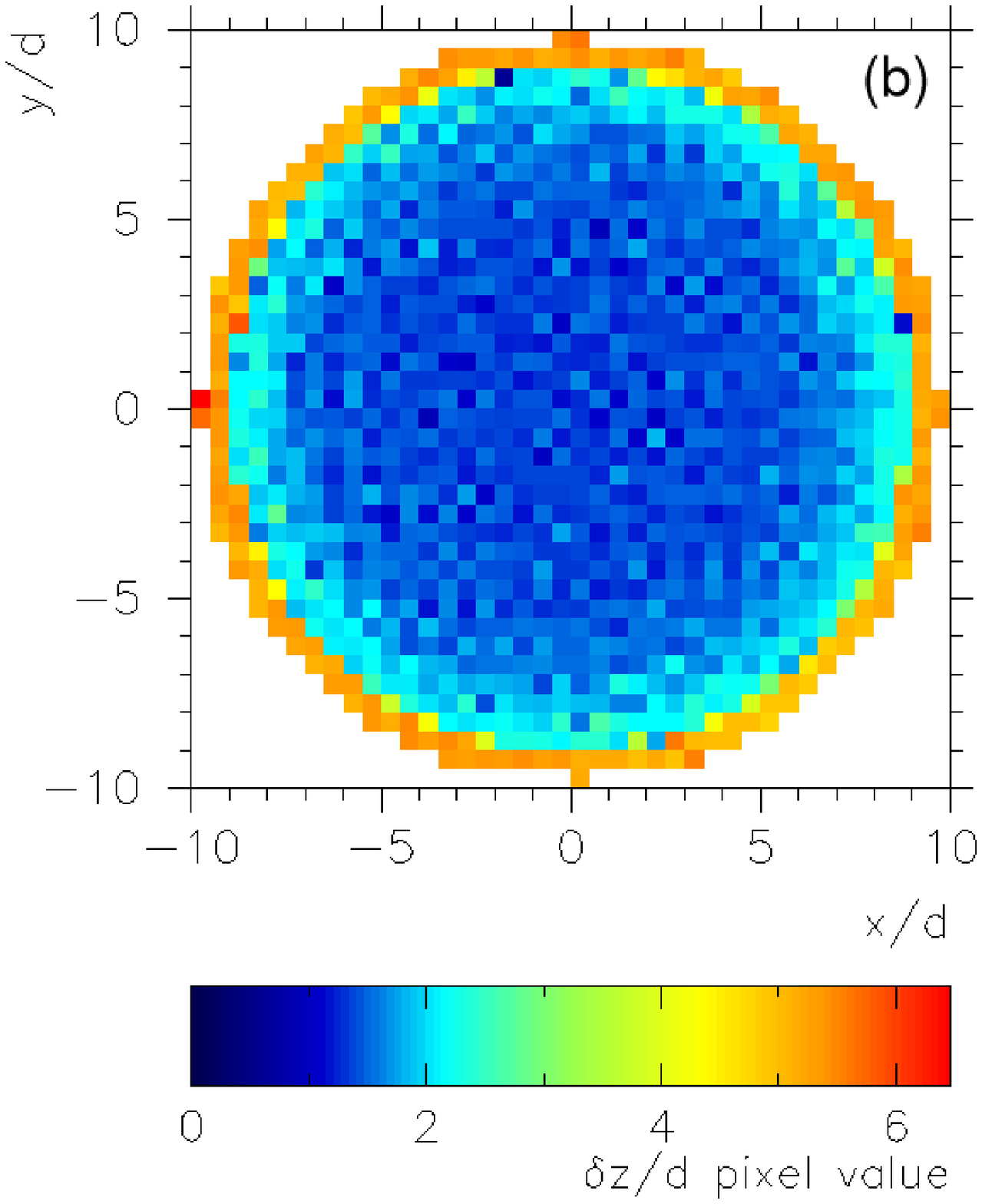}
\includegraphics[width=2.5in,clip]{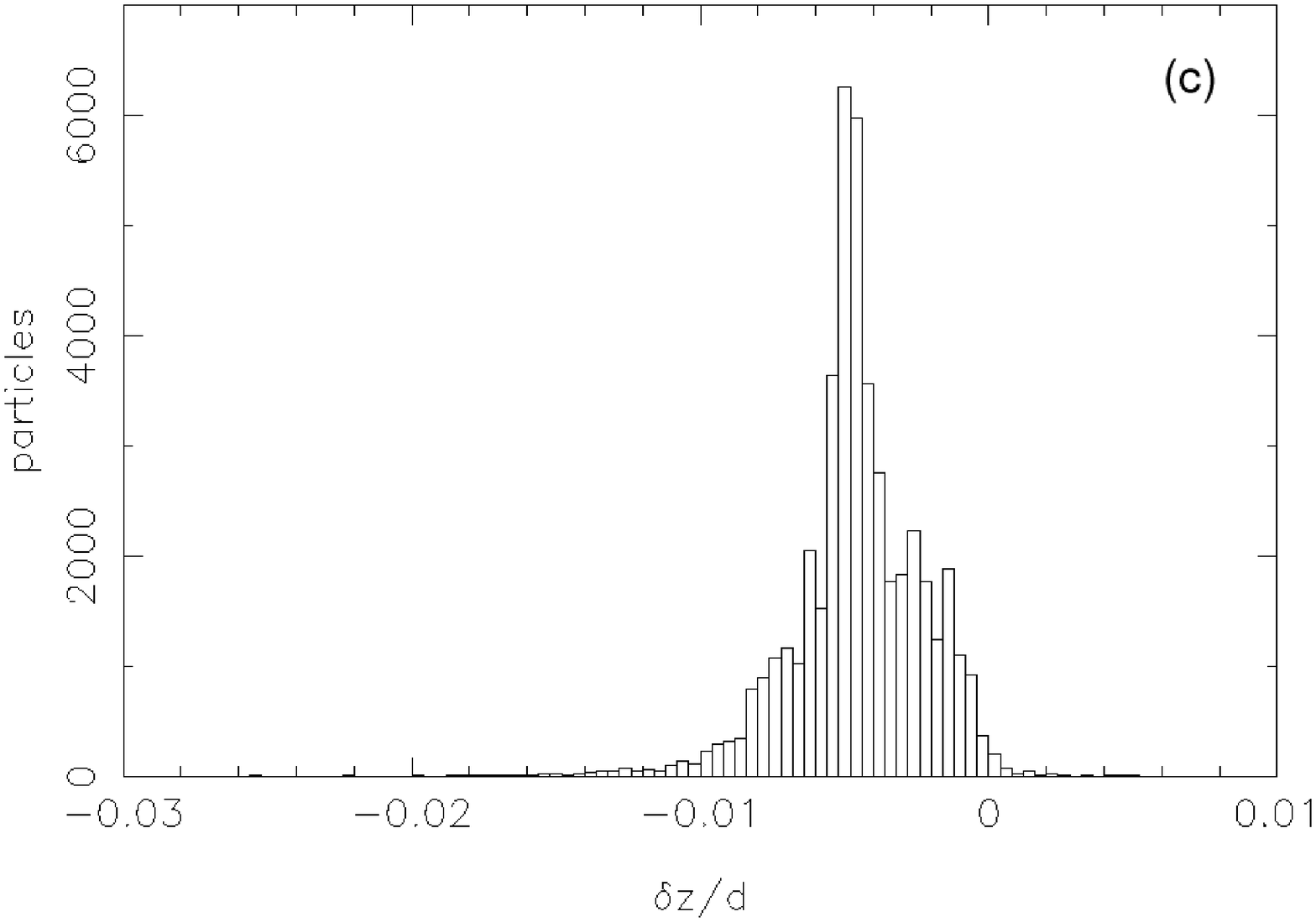}
\includegraphics[width=2in,clip]{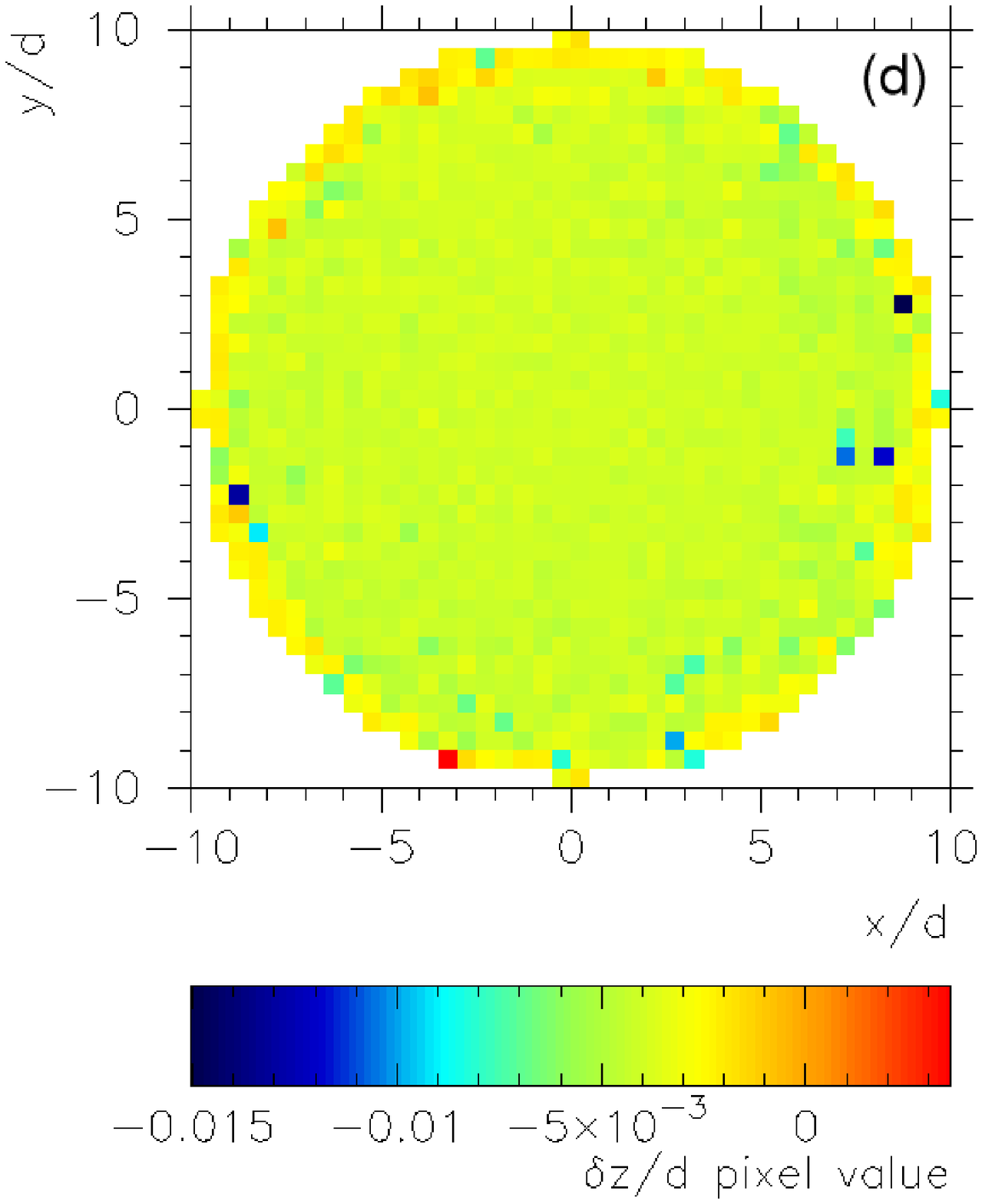}
\caption{\label{fig:dz-fast} $\delta z$, distance traveled by particles
  in the $z$ direction for $v_s = 10^{-2}\,d/\tau$.  (a) Histogram of
  $\delta z/d$ starting from static packing and moving walls for
  $t=10^{3}\tau$.  Particles have moved upward up to $7d$.  (b) The same
  data averaged over $z$ and presented as an image map.  There is only a
  small region of rapid upwards movement near the wall.  The bulk of the
  sample moves upward only slowly.  The following two plots cover the
  changes in $\delta z/d$ after wall movement has ceased and the packing
  has settled.  (c) Histogram of $\delta z/d$ from cessation of wall
  movement to a completely relaxed state.  Particles settle, but not
  nearly as much as the particles traveled over the course of the wall
  movement.  (d) same image map as in (b).  Here the particles settle
  more in the center than on the sides.}
\end{figure}

We present two sets of data for both high and low velocities.  The first
data set is for a high wall velocity of $v_s = 10^{-2} d/\tau$, and is
presented in two ways.  The first is a histogram of the motion of the
particles in $z$ from their initial starting position to the final
position after wall movement for $t = 10^3 \tau$, $\Delta z = 10d$.  The
measured movement of the particles in the $z$-direction $\delta z$ is
recorded and a histogram is generated.  Only those particles with an
initial position $z \ge 10d$ are included.  In the second method we
average $\delta z$ over $z$ and present an image map of the motion of
these particles in $z$.  Both the top ($z > 120d$) and bottom ($z <
10d$) regions of the initial packing are excluded to avoid edge effects.
These data sets are presented for the packing after wall movement for $t
= 10^3 \tau$ and then after a relaxation to equilibrium in
Figure~\ref{fig:dz-fast}.

For high wall velocities, there is significant particle movement.  In
this case, particles travel upwards as much as $7d$.  However, all of
this movement is restricted to a small ring around the cylinder walls.
In the center of the pack, particle movement is less than $d$.
Particles slide past one another and the initial stress networks
are quickly destroyed.  New force networks are constantly created and
destroyed over the course of the wall movement.

The initial drop in the saturation stress when the side wall has moved
approximately $0.3d$ is not enough to move the particles past each
other.  This minimum saturation stress thus reflects a particle
configuration where the initial contacts have been strengthened.  Once
particles at the wall move past their original contacts, the original
stress network is completely disrupted, and later networks as they form
and dissolve support less stress, causing the saturation stress to
increase.

When the wall movement stops, the packing settles, and the new $\delta
z$ is much smaller in magnitude than the $\delta z$ during wall
movement.  Also, the particles at the wall move much less than the
particles in the center, which suggests that the particles at the wall
are supported by friction at the walls.

\begin{figure}
\includegraphics[width=2.5in,clip]{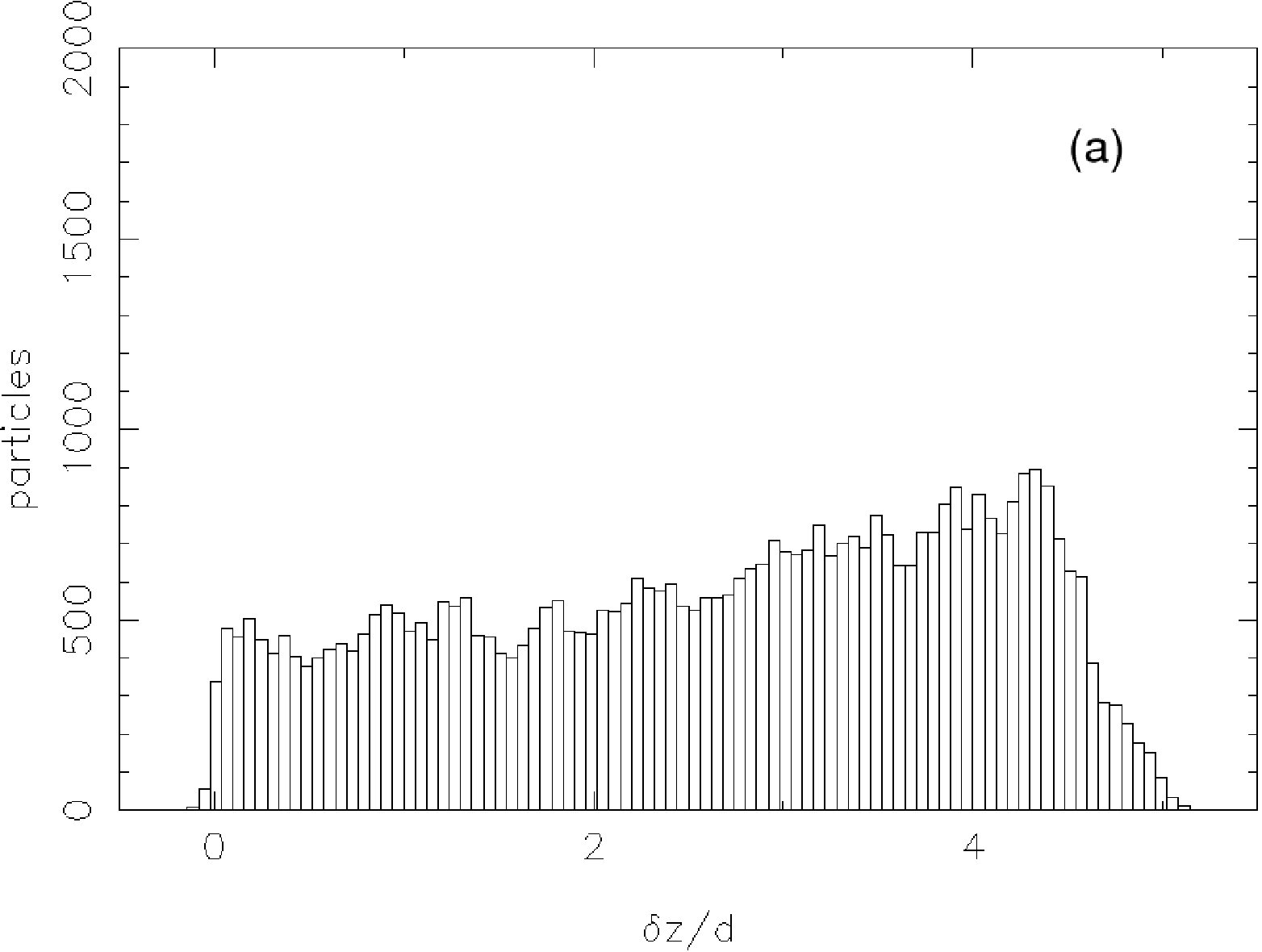}
\includegraphics[width=2in,clip]{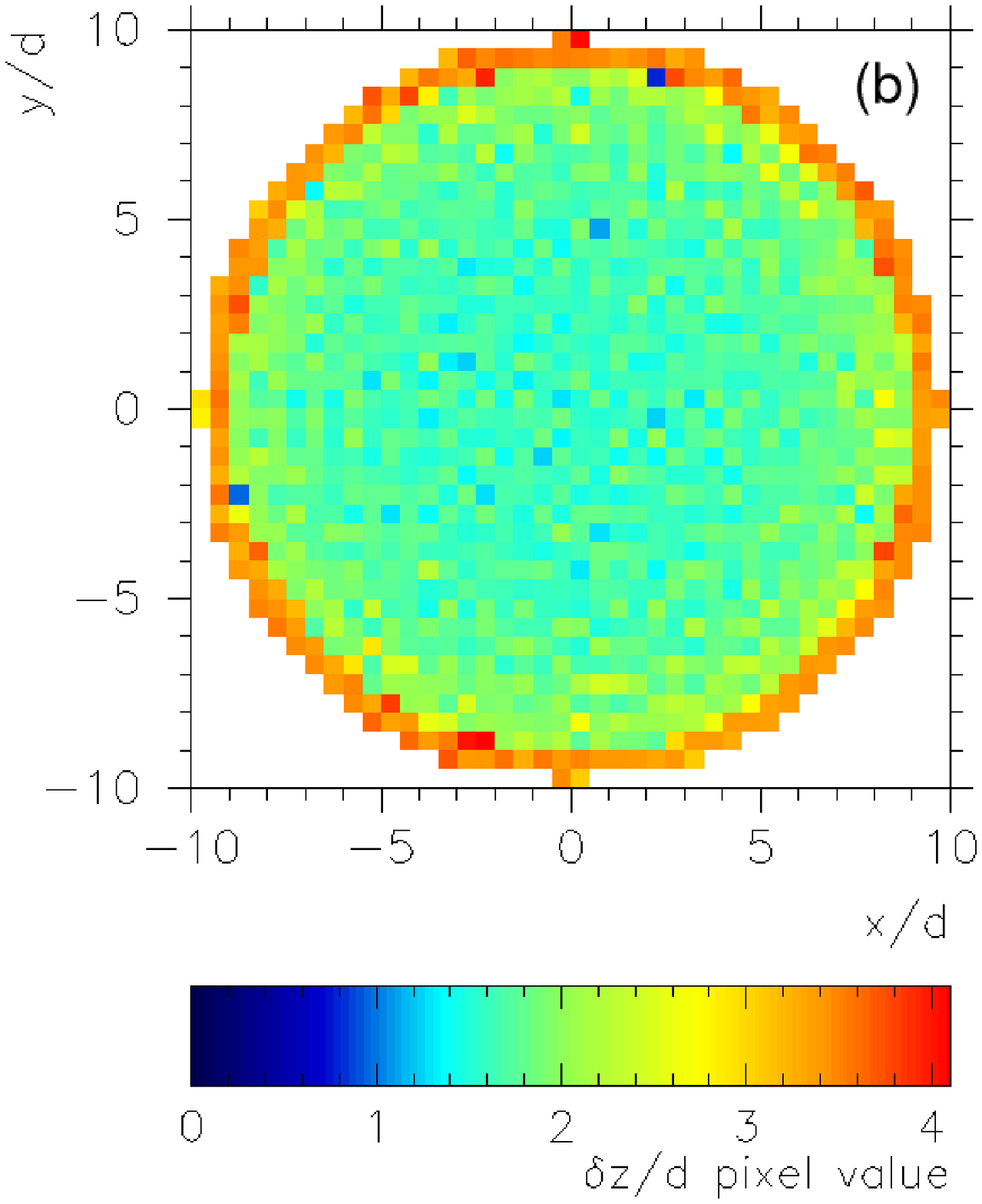}
\includegraphics[width=2.5in,clip]{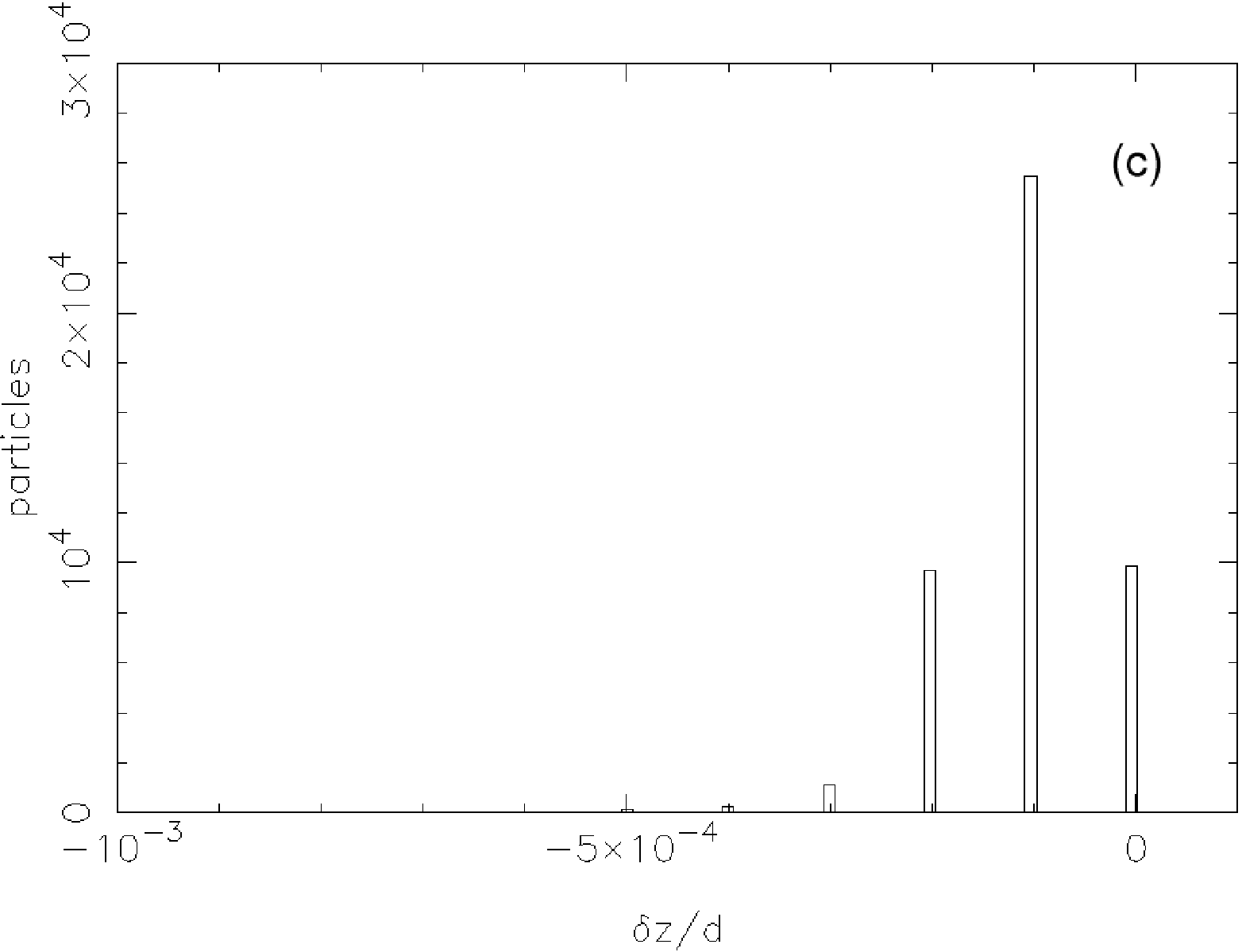}
\includegraphics[width=2in,clip]{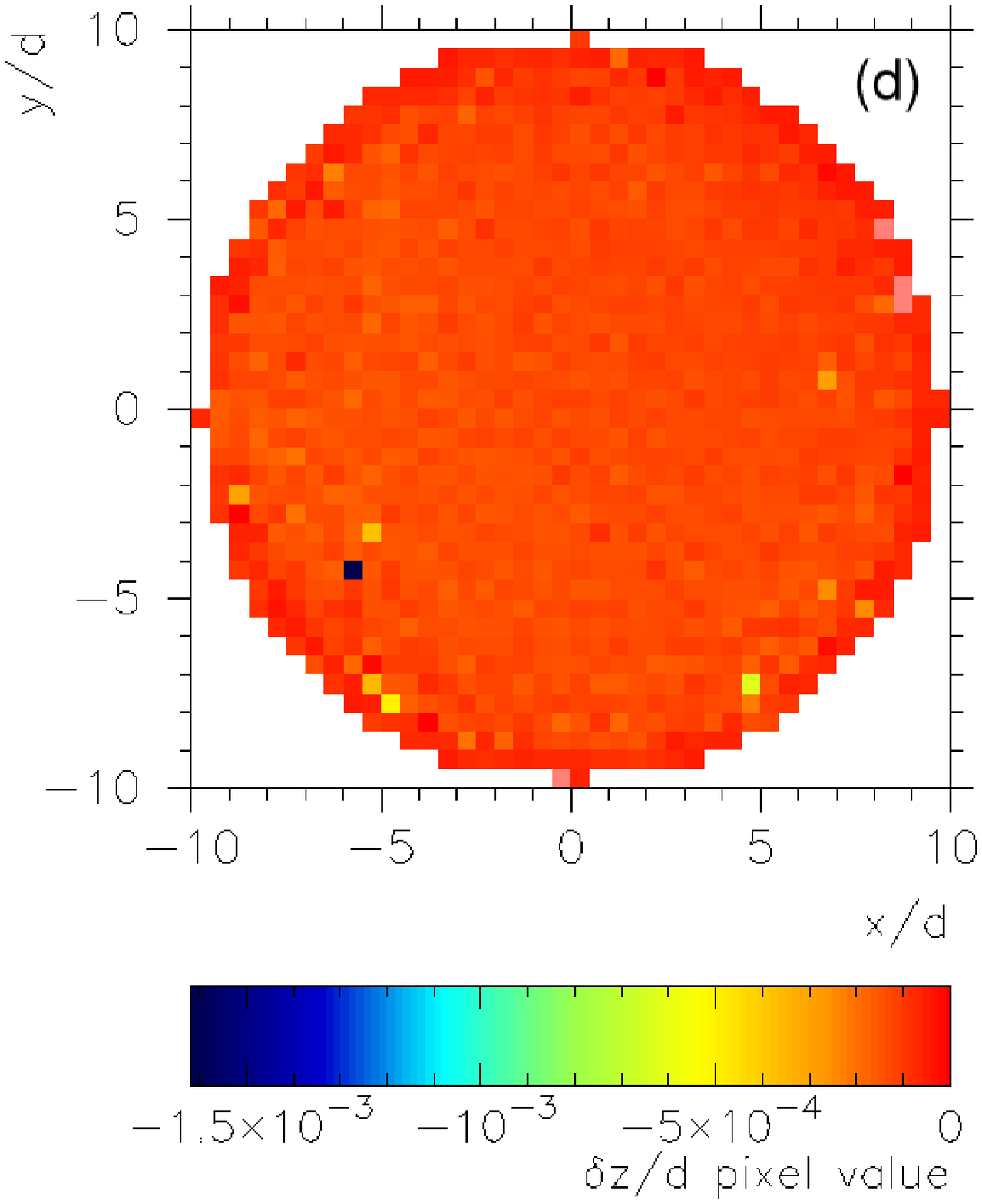}
\caption{\label{fig:dz-long} $\delta z$, distance traveled by particles
  in the $z$ direction over time for $v_s = 5 \times 10^{-4} d/\tau$.
  (a) Histogram of $\delta z/d$ starting from static piles and moving
  walls for $t = 1.14 \times 10^{4} \tau$.  Particles have moved up more
  than $5d$ as the walls have moved $\Delta z = 5.7d$.  (b) The same
  data averaged over $z$ and presented as an image map.  There is only a
  small layer of rapid upwards movement.  The bulk of the sample moves
  upward only slowly.  (c) Histogram of $\delta z/d$ from cessation of
  wall movement to a completely relaxed state.  Particles settle, but
  not nearly as much as the particles traveled over the course of the
  wall movement.  (d) same image map as in (b).  Here the particles
  settle more in the center than on the sides, though the difference is
  not as great as in the high velocity case.}
\end{figure}

Figure~\ref{fig:dz-long} shows $\delta z$ for the low wall velocity
$v_s = 5 \times 10^{-4} d/\tau$ applied for the much longer time $t =
1.14 \times 10^4 \tau$ for a total wall movement of $\Delta z = 5.7d$.
The change in particle heights for $\delta z$ is similar to that seen in
high velocity runs.  Particles along the walls move upward much more
than those in the center.  However, unlike in the high-velocity case,
there is no second ring of particles one or two $d$ in from the wall
that also shows a large $\delta z$ relative to the center.  In this
case, the particles not in contact with the wall show much less
difference in $\delta z$ as a function of radius.

Relaxation is very interesting in this case.  The difference in $\delta
z$ between particles at the wall and particles in the bulk is much less
than in the high velocity case.  In addition, relaxation occurs in
discrete jumps with many particles moving together.  This is a general
feature of $\delta z$ relaxation histograms for low wall velocity
packings.  In the low velocity case, energy is imparted into the system
very slowly and is dissipated more quickly.  There is thus much less
energy available to rearrange the packing after the wall movement
ceases, and the relaxation of the particles is much less.  In addition,
because individual particles have very little kinetic energy when the
wall movement ceases, motion happens coherently as many particles
rearrange at once.

\section{Coulomb Criterion}

\begin{figure}
\includegraphics[width=2.25in,angle=270,clip]{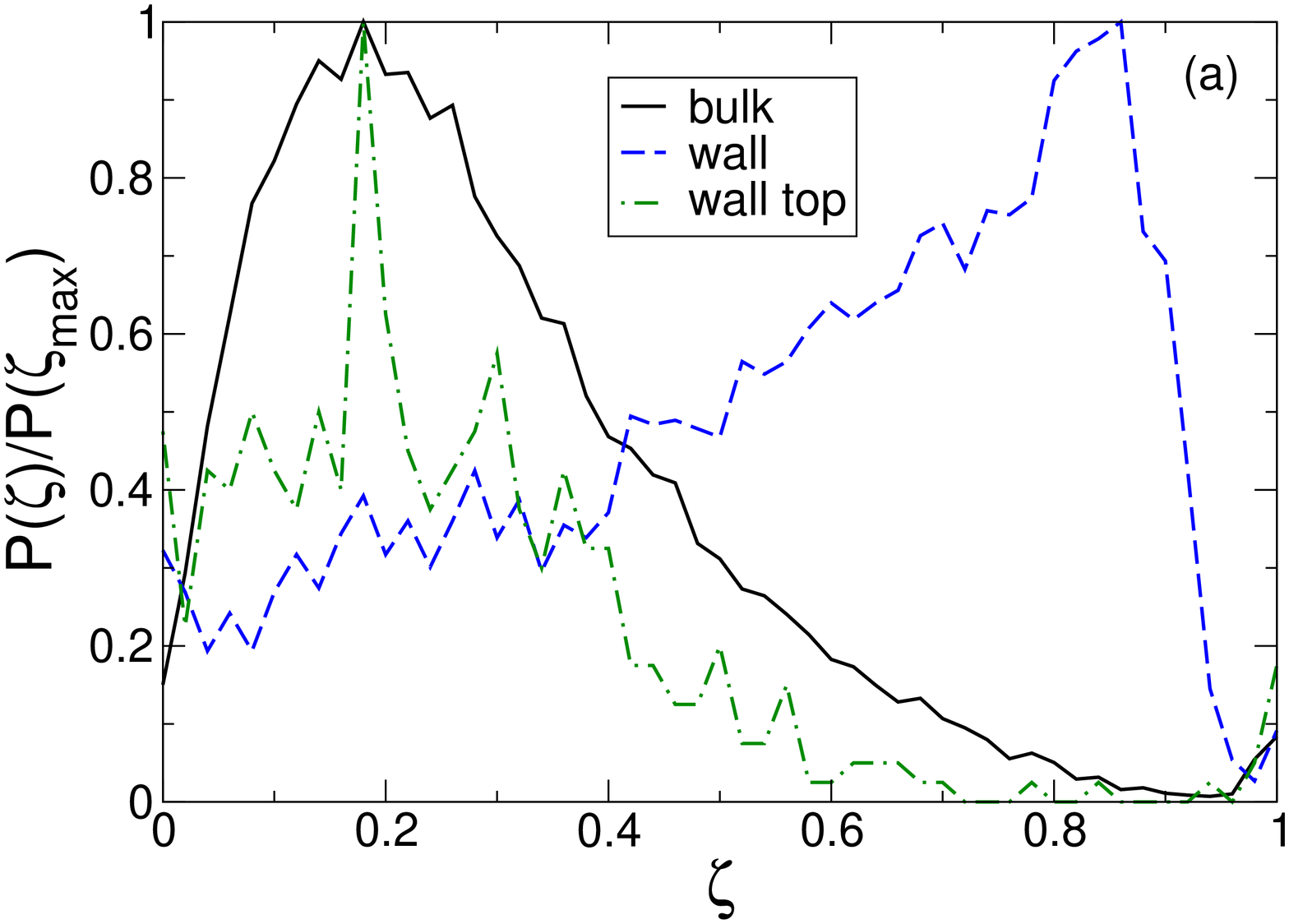}
\includegraphics[width=2.25in,angle=270,clip]{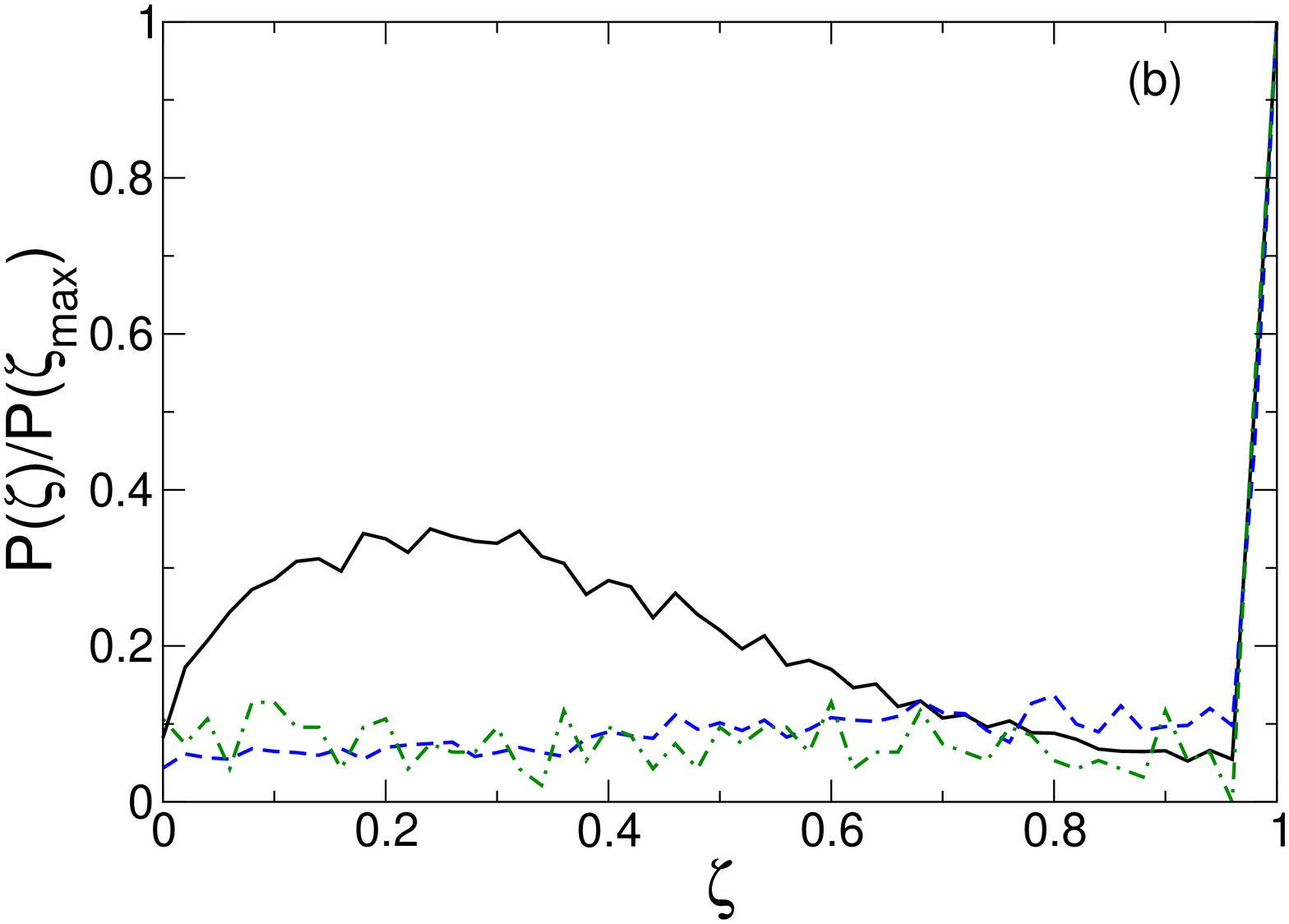}
\includegraphics[width=2.25in,angle=270,clip]{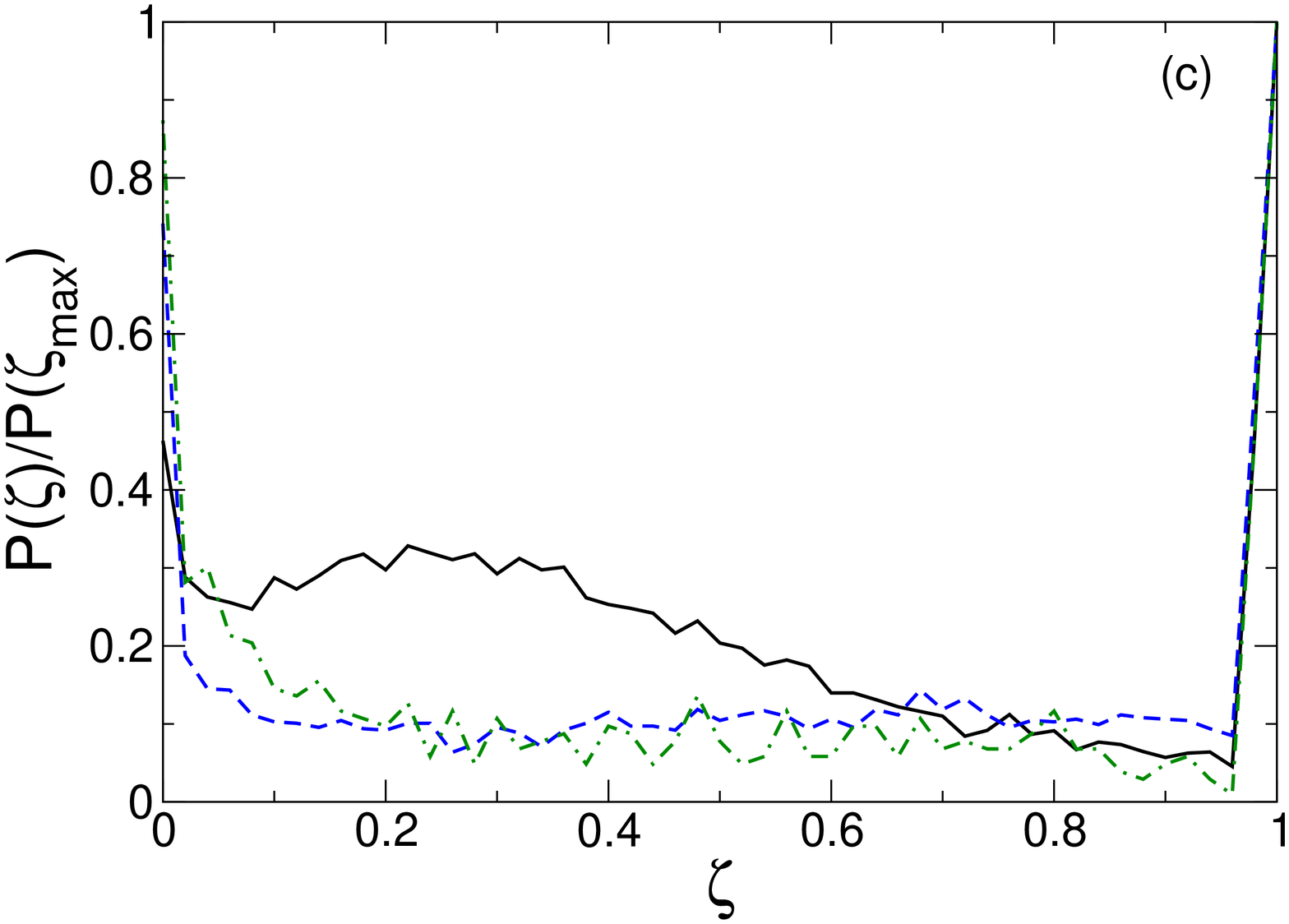}
\includegraphics[width=2.25in,angle=270,clip]{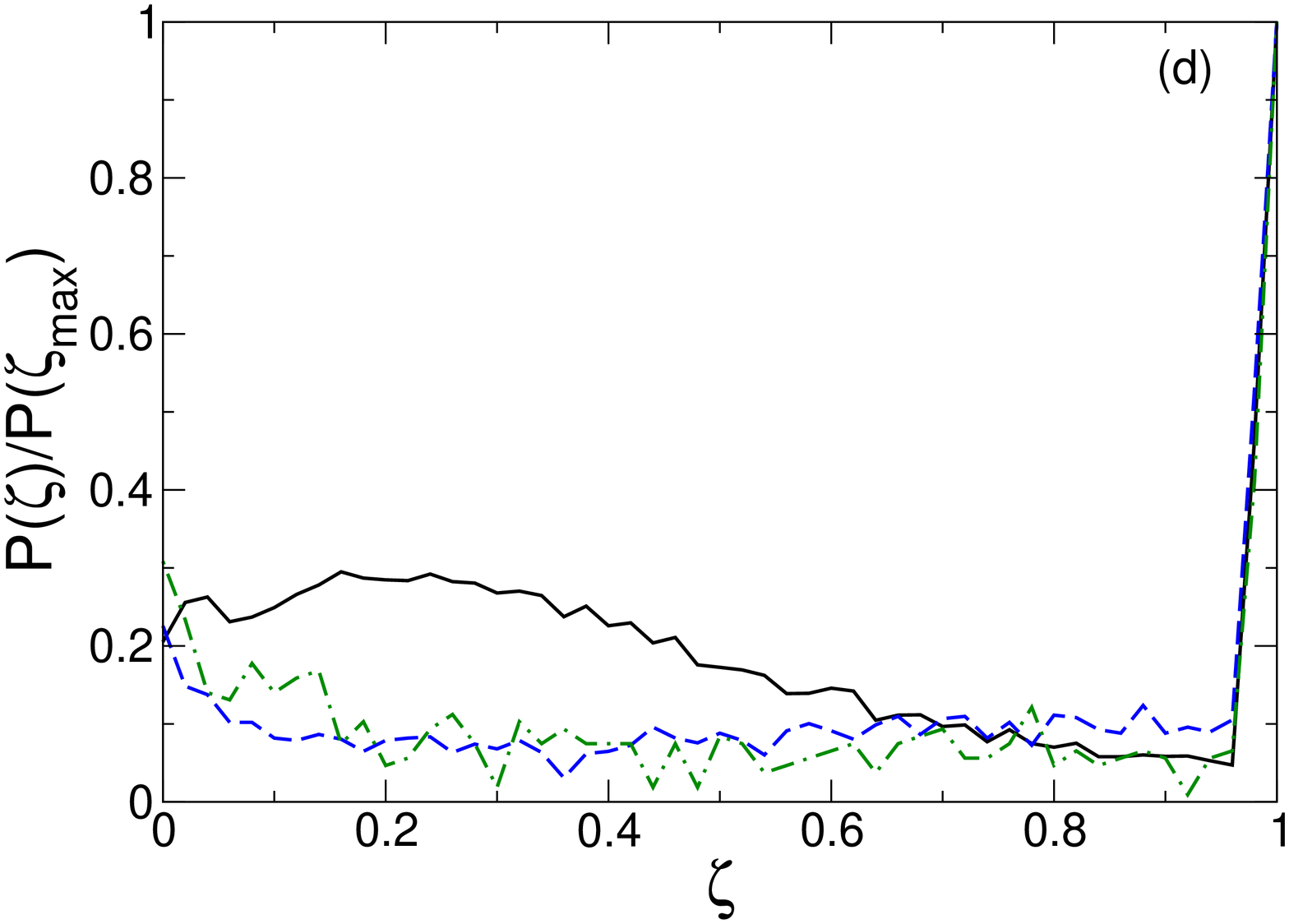}
\caption{\label{fig:tvn} Ratio of normal forces to tangential forces in
  packings.  The force distributions were analyzed in three different
  locations: bulk - in the depths of the packing away from the walls
  (solid line), wall - only the particle-wall interactions in the center
  of the pile (dashed line), and wall top - only the particle-wall
  interactions near the top of the pile (dot-dashed line).  $\zeta =
  F_t/ \mu F_n$ in the case of the bulk, $\zeta = F_t/ \mu_w F_n$
  otherwise.  (a) Original quiescent packing.  (b) After $v_s =
  10^{-2}\,d/\tau$ for $t = 10^{3} \tau$, $\Delta z = 5d$.  (c) Same as
  in (b) after cessation of wall movement and relaxation (d) After $v_s
  = 5 \times 10^{-4}\,d/\tau$ for $t = 10^5 \tau$, $\Delta z = 5d$.}
\end{figure}

We also examined the force distributions in the packings during and
after wall movement.  These provide further evidence that the final
packing state after relaxation is very different from the initial state,
and also demonstrate the different behavior of the two regimes after
relaxation.  A useful quantity is the ratio of normal forces $F_n$ to
tangential forces $F_t$.  A key assumption of the Janssen analysis is
that the ratio of these forces $\zeta = F_t/ \mu_w F_n = 1$ for
particle-wall forces in the packing.  These distributions are shown in
Figure~\ref{fig:tvn} for several different packings.  We consider three
different distributions: forces in the bulk of the sample far from the
wall, particle-wall forces in center of the packing, and particle-wall
forces at the top of the packing. In this case, $\zeta = F_t/ \mu F_n$
in the bulk of the sample, and $\zeta = F_t/ \mu_w F_n$ for
particle-wall forces.

The original packing has a strong peak near $\zeta = 0.2$ in the bulk of
the material, showing that the vast majority of forces are far from the
Coulomb criterion.  At the wall, the peak in the distribution is much
closer to the Coulomb criterion at $\zeta \sim 0.9$.  At the top of the
packing, in the hydrostatic region~\cite{LandryApr2003}, the
distribution of forces is far from the Coulomb criterion.  

After wall movement, the distribution of forces is radically changed.
For both the high velocity ($v_s = 10^{-2}\,d/\tau$) and low velocity
($v_s = 5 \times 10^{-4}\,d/\tau$) cases, all three distributions are
driven toward the Coulomb criterion, with the peak in the distributions
at the Coulomb criterion.  In both cases, the original distribution is
not completely destroyed. For the bulk distribution, a subsidiary peak
appears in the new bulk distribution where the original peak appeared.
This change in the distribution of forces is the main cause for the more
Janssen-like stress distributions observed.

We also present results for the $v_s = 10^{-2}\,d/\tau$ case after
cessation of wall movement and relaxation in Fig.~\ref{fig:tvn}c.  In
this case, the main change in the distribution after relaxation is the
increase in particles forces at very low $\zeta$.  This occurs because
some contacts between particles disappear during relaxation as particles
move relative to their neighbors.  When new contacts are made, these are
far from the Coulomb criterion by definition, since particles are not,
in general, rotating relative to each other during relaxation.  By
contrast, the low velocity case $v_s = 5 \times 10^{-4}\,d/\tau$,
Fig.~\ref{fig:tvn} exhibits no change in the distribution of forces after
cessation and relaxation.  In this case, particles move together, and
contacts are not destroyed or created, so the force distribution does
not change.  This lack of change in the force distribution may explain
why low velocity packings do not exhibit perfect agreement with the
Janssen form after relaxation.

\section{\label{sec:conclusion}Conclusions}

We have explored the effects of side wall movement on granular packings.
Much of the resultant structure of the pack does not depend strongly on
the magnitude of the wall movement, as long as the packing is moved for
an equivalent distance.  Small differences emerge in the behavior after
cessation, where high wall velocity packings substantially rearrange
and increase their saturation stress while low wall velocity packings
remain essentially unchanged.  The main effect of wall movement is to
drive both the particle-wall and particle-particle contacts to the
Coulomb criterion, so that the ratio of tangential forces to normal
forces is maximized.  This condition forces the packing in the high wall
velocity case to obey the Janssen form, which takes the Coulomb
criterion as one of its main assumptions.  For low wall velocities, the
final form is very different from the initial form, though it does not
perfectly match the Janssen form.

This work was supported by the Division of Materials Science and
Engineering, Basic Energy Sciences, Office of Science, U.S. Department
of Energy.  This collaboration was performed under the auspices of the
DOE Center of Excellence for the Synthesis and Processing of Advanced
Materials.  Sandia is a multiprogram laboratory operated by Sandia
Corporation, a Lockheed Martin Company, for the United States Department
of Energy's National Nuclear Security Administration under contract
DE-AC04-94AL85000.

\bibliography{grain}

\begin{thebibliography}{19}
\bibitem[{Her()}]{HertzianNote1}
\bibinfo{note}{We have made a packing using Hertzian springs and analyzed its
  behavior under wall movement and found no substantial difference in its
  behavior relative to that presented here.}
\end{thebibliography}

\end{document}